\documentclass[aps,prd,reprint,twocolumn,nofootinbib]{revtex4-2}
\usepackage{amsmath}
\usepackage{amssymb}
\usepackage{verbatim,graphics,graphicx,color,slashed,textcomp,bbm,mathdots,multirow,array}
\usepackage[normalem]{ulem} 
\usepackage[colorlinks=true,linkcolor=red,urlcolor=blue,citecolor=blue]{hyperref}

\usepackage{autobreak}
\usepackage{float}

\graphicspath{{figures/}}

\usepackage{xcolor}

\begin{document}
	
\title{\bf Signatures of Ultralight Bosons in Compact Binary Inspiral and Outspiral}

\author{Yan Cao$^{a}$}
\author{Yong Tang$^{a,b,c,d}$}
\affiliation{\begin{footnotesize}
		${}^a$School of Astronomy and Space Sciences, University of Chinese Academy of Sciences (UCAS), Beijing 100049, China\\
		${}^b$School of Fundamental Physics and Mathematical Sciences, \\
		Hangzhou Institute for Advanced Study, UCAS, Hangzhou 310024, China \\
		${}^c$International Center for Theoretical Physics Asia-Pacific, Beijing/Hangzhou, China \\
		${}^d$National Astronomical Observatories, Chinese Academy of Sciences, Beijing 100101, China
		\end{footnotesize}}

\date{\today}
\begin{abstract}
Ultralight bosons are well-motivated particles from various physical and cosmological theories and can be spontaneously produced during the superradiant process, forming a dense hydrogenlike cloud around the spinning black hole. After the growth saturates, the cloud slowly depletes its mass through gravitational-wave emission. In this work we study the orbit dynamics of a circular binary system containing such a gravitational atom saturated in various spin-0, -1 and -2 superradiant states, taking into account both the effects of dynamical friction and the cloud mass depletion. We estimate the significance of mass depletion, finding that although dynamical friction could dominate the inspiral phase, it typically does not affect the outspiral phase driven by the mass depletion. Focusing on the large orbit radius, we investigate the condition to observe the outspiral and the detectability of the cloud via pulsar-timing signal in the case of black hole–pulsar binary.
\end{abstract}

\maketitle


\newpage
\section{Introduction}
Ultralight bosons are well-motivated particles from various theories beyond the Standard Model and can be good candidates of dark matter (DM)~\cite{Peccei.Quinn, Wilczek.1978, PhysRevLett.40.223, Preskill:1982cy, Abbott:1982af, Dine:1982ah, Svrcek:2006yi, Arvanitaki_2010, Graham:2015rva, Ema:2019yrd, Ahmed:2020fhc, Hui:2016ltb, Hui:2021tkt, Ferreira_2021, 2168507, Antypas:2022asj, Tang:2020ovf, Kitajima:2023fun, Kolb:2020fwh, An:2020jmf, Moroi:2020has, Liang:2022gdk, Sun:2021yra, Sato:2022jya}. Their nongravitational couplings to normal matter are generally predicted to be extremely weak, so that experimental and astrophysical searches of these direct couplings can be rather difficult, and typically reply on the assumption of the boson's background abundance. The black hole (BH) superradiance (SR) \cite{book_superradiance, Arvanitaki:2010sy}, however, provides a natural astrophysical mechanism to produce these bosons solely from their minimum coupling to gravity. Due to the rotational superradiant instability in Kerr background, macroscopic condensate of free spin-0, -1 and -2 bosons can spontaneously develop around the host BHs by extracting their energy and angular momentum. The observational signatures of the resulted cloud-BH systems, so-called gravitational atoms (GAs), provide promising ways to detect these ultralight degrees of freedoms~\cite{PhysRevLett.119.131101, Hannuksela:2018izj, Zhu:2020tht, Yuan:2021ebu, Chen:2021lvo, Chen:2022kzv, Siemonsen:2022yyf, Siemonsen:2022ivj}.  
 
If the GA is part of a binary system, further interesting phenomenology arises already in the perturbative regime, such as the orbit-cloud resonances~\cite{B1,B2,B3,Berti_1,HSC}, dynamical friction (DF) or ionization~\cite{B4, B5, ZJ, Takahashi_2, Tomaselli:2023ysb}, companion-induced suppression of the SR instabilities~\cite{WY_3} and the cloud-induced orbit precession \cite{Su}. In discussion of these effects the gravity of the cloud is usually neglected (an exception being \cite{Ferreira_2017}, which studies the orbits of a small companion at relatively small distance from the cloud in scalar SR ground state), and the cloud mass is usually assumed to be not much smaller than its initial saturated value (an exception being~\cite{Takahashi:2023flk}). However, as first pointed out in \cite{Kavic:2019cgk} for the scalar GA, if the cloud mass is included in the orbit dynamics, the intrinsic mass depletion of the cloud (DC) due to its gravitational-wave (GW) emission would affect the orbit evolution in an opposite manner with other dissipative effects, i.e., it tends to make the binary outspiral, and this effect is actually important at large radius. Recently, the cloud mass depletion has also been considered in \cite{Xie:2022uvp} for the scalar cloud (albeit using a different sign for its effect), and in \cite{Fell:2023mtf} for a relativistic vector cloud in the SR ground state, but neglecting other cloud-induced dissipations. Besides the SR clouds, there are proposals to search for the anomalous orbit evolution due to mass variation arising, e.g., from enhanced BH evaporation~\cite{Simonetti:2010mk} and the accretion of background DM into the BH~\cite{Akil:2023kym}. 

In this work, we present a systematic model for the GA+companion system, describing various spin-0, -1 and -2 SR states in a unified manner. The focus is to study the interplay between binary GW emission, dynamical friction and the cloud mass depletion and to compare the situations for GA saturated in different SR states. The possibility of outspiral also has implications on the secular evolution of such systems and whether the binary could undergo fine and hyperfine resonances (taking place also at large radius). Finally, the cloud depletion may already leave imprints on the orbit evolution that is directly observable via GW and pulsar timing measurements, even in the absence of resonance events and dramatic mode mixings. Indeed, we find that for circular binary containing scalar and vector atoms in their SR ground states, the observable parameter space is largely independent from the dynamical friction, though the inclusion of DF enlarges the observable regions.

The structure of this paper is as follows: In Sec.~\ref{sec_2} we review the properties of an isolated GA and formulate the binary model. Then in Sec.~\ref{sec_3} we discuss the orbit evolution quantitatively, and in Sec.~\ref{sec_4} investigate the detectability of the cloud from orbit phase measurements. Finally, we summarize our results and discuss some possible future directions in Sec.~\ref{sec_5}. Throughout our discussion, unless stated otherwise, we use the natural unit with $\hbar=G=c=1$.

\section{Gravitational Atom in Binary}\label{sec_2}
\subsection{Gravitational Atom}
First we briefly summarize the main properties of the nonrelativistic superradiant clouds around Kerr BH. For the physical (spatial) components of a real bosonic field $\boldsymbol{\Phi}$, far away from a central mass $M$, the wave function $\boldsymbol{\Psi}$ defined by $\boldsymbol{\Phi}=\frac{1}{\sqrt{2\mu}}(\boldsymbol{\Psi} e^{-i\mu t}+\text{c.c.})$ (and $\frac{1}{\sqrt{\mu}}\boldsymbol{\Psi} e^{-i\mu t}$ if $\boldsymbol{\Phi}$ is complex, but we shall focus on the real fields) satisfies the Schrodinger equation
\begin{equation}
	i\partial_t\boldsymbol{\Psi}=-\frac{1}{2\mu}\nabla^2\boldsymbol{\Psi}-\frac{\alpha}{r}\boldsymbol{\Psi} ~,
\end{equation}
where $\mu$ is the mass of the boson and $\alpha\equiv \frac{GM\mu}{\hbar c}=M\mu$ the gravitational fine structure constant. For scalar fields $\boldsymbol{\Psi}=\psi$, for Proca fields $[\boldsymbol{\Psi}]_i=\psi_i$ and for spin-2 tensor fields $[\boldsymbol{\Psi}]_{ij}=\psi_{ij}$. This is as the same as the Schr\"{o}dinger equation for hydrogen atom (for each field component) with well-known bound state solutions; in case that the central body is a Kerr BH, these hydrogenic states can be spontaneously populated by rotational superradiance and a GA is formed. In this work we focus on GA with $\alpha\ll 1$ (hence the Bohr radius $r_c=M/\alpha^2 \gg M$), for which this nonrelativistic Newtonian description is appropriate. The mass density of the cloud (same for both real and complex $\boldsymbol{\Phi}$ fields) is given by $\rho=M_c\text{Tr}(\boldsymbol{\Psi}^\dagger\boldsymbol{\Psi})$, where we choose the normalization $\int d^3x \rho=M_c$ and $M_c$ is the total mass of the cloud. For convenience we also define $\beta\equiv M_c/M$.

The cloud is generally a superposition of all bound atomic levels, $|\boldsymbol{\Psi}\rangle=\sum_ic_i|\boldsymbol{\Psi}_i\rangle$. Then using an orthonormal basis $\langle \boldsymbol{\Psi}_i|\boldsymbol{\Psi}_{i'}\rangle=\delta_{ii'}$, we have $\sum_i|c_i|^2=1$. However, the modes are not static. In the case of a single occupied mode this time dependence can be absorbed to $M_c(t)$. For multiple modes this is not feasible, and it is more convenient to track the evolution of individual $c_i$.

The eigenstates are labeled by the quantum numbers $n,l,j$ and $m$, the principal, orbit angular momentum, total angular momentum and azimuthal quantum number, respectively (for scalar GA, $j=l$, so we write $n,l,m$), corresponding to an orthonormalized wave function $|nljm\rangle = \boldsymbol{\Psi}_{nljm}(t,\mathbf{r})$ (the detailed forms are listed in Appendix~\ref{Appendix_A}). For a spin-s field, the quantum numbers satisfy $n\ge 1, l\in [0,n-1], j\in [|l-s|,l+s]$ and $m\in[-j,j]$. The real part of the energy level $\omega\equiv \omega_R+i\omega_I$ is given by $\mu(1-\frac{\alpha^2}{2n^2}+\mathcal{O}(\alpha^4))$. Crucially, $\omega$ also contains an imaginary part, and the superradiant growth can occur only if $\omega_I>0$, i.e., when $\omega_R/m<\Omega_H\equiv \frac{1}{2M}\frac{\chi}{1+\sqrt{1-\chi^2}}$, which demands a large enough BH spin $\chi$. Starting with a sufficiently fast-spinning bare black hole, the superradiant growth is expected to be dominated by the most unstable mode, which is the 211 state for scalar GA, the 1011 state for vector GA and the 1022 state for spin-2 GA \footnote{The situation is a little more complicated for spin-2 atom \cite{Brito_2020}, the 1022 and 2111 states grow simultaneously, yet by the time 1022 state saturates with $\Omega_H$ spinning down to $\omega_R/2$, the 2111 state has been completely reabsorbed. For comparison, we shall also include the possibility of a saturated 2111 state. Also, for $\alpha\sim\mathcal{O}(0.1)$, the fastest growing spin-2 state is a non-hydrogenic dipole mode with $m=1$ \cite{Dias:2023ynv}. But here we are interested in the regime $\alpha\ll 1$.}. The cloud then slowly decays by its intrinsic gravitational-wave emission after the instability saturates, until the growth of the next unstable mode \cite{B3}, which is the 322 state for scalar GA and the 2122 state for vector GA. Depending on the initial condition, the cloud may also be occupied by multiple modes \cite{Ficarra:2018rfu}. In this work we shall focus on a single saturated mode, that would be characterized by its mass distribution $\rho(\mathbf{r})$ and mass depletion rate $P_{gw,c}$. Generically, the density profile and depletion power have the scaling form 
\begin{equation}
\rho(\mathbf{r})=\frac{M_c}{r_c^3}g(x,\theta) ~,
\end{equation}
with $x\equiv r/r_c$, and
\begin{equation}
P_{gw,c}=\beta^2 p(\alpha) ~,
\end{equation}
where the dimensionless function $g(x,\theta)$ and $p(\alpha)$ are state dependent. We use the accurate polynomial fits of $p(\alpha)$ provided in \cite{Siemonsen:2022} for scalar and vector states and the analytical approximation for tensor states with $\alpha\ll 1$ calculated in \cite{Brito_2020}. In physical units,
\begin{equation}
\rho(x)=g(x,\theta)\,\frac{\beta}{0.1}\left(\frac{\alpha}{0.1}\right)^6\left(\frac{M_\odot}{M}\right)^2\times 3.46\times 10^{34}\,\text{GeV}/\text{cm}^3.
\end{equation}
The functions $g(x,\pi/2)$ for various SR states are plotted in Fig.~\ref{density} and their complete expressions are listed in Appendix~\ref{Appendix_A}. The interested bosonic field has a typical mass
\begin{equation}
\mu= \left(\frac{\alpha}{0.1}\right)\left(\frac{M_\odot}{M}\right)\times1.3\times 10^{-11}~\text{eV}.
\end{equation}

Neglecting the change of black hole mass, we have $\dot M_c=2\omega_I M_c=M_c/\tau_I$ during the superradiant growth. Thus the cloud mass grows exponentially in an instability timescale $\tau_I\equiv 1/2\omega_I$. When the mass change of the cloud is dominated by GW emission, $-\dot M_c=P_{gw,c}\propto M_c^2$, the cloud mass decays according to
\begin{equation}
M_c(t)=\frac{M_{c,0}}{1+\frac{t-t_0}{\tau_{gw}}} ~,
\end{equation}
where $\tau_{gw}\equiv \frac{M_{c,0}}{P_{gw,c}}=\frac{M}{\beta p(\alpha)}$ is the mass depletion timescale. For $\alpha\ll 1$, $\tau_{gw}\gg \tau_I$, the GW emission can be neglected in the superradiant growth. From energy and angular momentum conservation, the mass or angular momentum  of the cloud after the growth (of a mode with azimuthal quantum number $m$) saturates is given by the difference between initial and final BH mass or spin: 
\begin{equation}
M_{c,0}=M_i- M_f,\quad J_f=J_i-\frac{m}{\omega_R}(M_i-M_f) ~,
\end{equation}
with \cite{Tsukada:2018mbp} 
\begin{equation}
M_f=\frac{m^3-\sqrt{m^6-16 m^2 \omega_R^2\left(m M_i-\omega_R J_i\right)^2}}{8 \omega_R^2\left(m M_i-\omega_R J_i\right)} ~.
\end{equation}
For $\chi_i\approx 1$, $M_{c,0}\approx \frac{\alpha}{m} M_i$ and the saturated BH spin is $\chi_{f}\approx \frac{4\alpha}{m}$. The spin can be further extracted by the next growing mode $m'$. Then the saturated values become $M_{c,0}\approx \frac{4(m'-m)}{mm'^2}\alpha^2 M_i$ and $\chi_f\approx \frac{4\alpha}{m'}$.

\begin{figure}[t]
	\centering
	\includegraphics[width=0.45\textwidth]{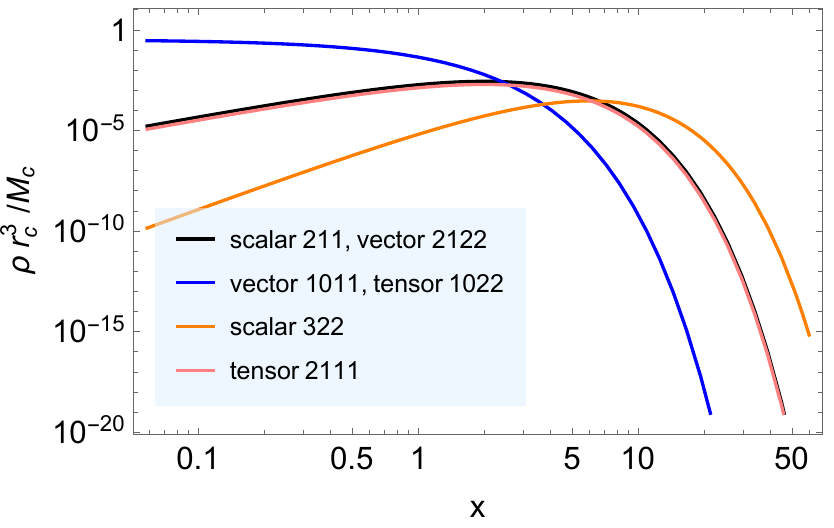}
	\caption{The density distribution of spin-0, -1, -2 superradiant ground states on the equatorial plane.}\label{density}
\end{figure} 

\subsection{Binary Orbit Dynamics}
Now we consider the situation when the saturated GA belongs to a binary system\footnote{In principle, the companion can also carry its own environment, but we neglect this since the companion is assumed to be small.} and contrast the various possible effects induced by the cloud. We focus on the Keplerian circular orbits on the equatorial plane of the host BH (for a brief discussion of inclined and elliptical orbits see Appendix~\ref{Appendix_C}) and assume a large orbit radius so that the cloud's tidal distortion is completely negligible. As we shall see, the binary motion can still receive significant modifications due to the presence of the cloud.

We take the BH mass $M$ and the companion's mass $M_*\equiv qM$ to be constant, the Newtonian orbit energy and angular momentum are given, respectively, by
\begin{equation}
E=-\frac{(M+\tilde M_c)M_*}{2r}
~,\quad
L=
\sqrt{\frac{((M+\tilde M_c)M_*)^2}{M+ \tilde M_c+M_*}r}~,
\end{equation}
where $\tilde M_c(\mathbf{r})$ is the effective cloud mass experienced by the companion (the detailed definition is given in Appendix~\ref{Appendix_B}). In the following we shall take the limit $M+\tilde M_c\approx M$. To restore $\tilde M_c$ one needs just the replacement $M\to M+\tilde M_c$ and $q\to \frac{M}{M+\tilde M_c}q$; however, such corrections remain small and do not affect the main results. Without the mass change, the orbit would evolve according to
\begin{equation}
-\dot E=P_{gw}+P_{others}~,
\end{equation}
where $P_{gw}$ is the binary GW radiation power:
\begin{equation}
P_{gw}(x)=\frac{32}{5}\frac{\alpha^{10} (1+q)q^2}{x^5}~,
\end{equation}
(the correction to this power due to cloud depletion is negligible, see Appendix~\ref{Appendix_B}) and $P_{others}$ is the contribution from extra dissipation channels. In the Newtonian order, the orbit evolution can also be written as 
\begin{equation}\label{orbit}
\dot x=-\frac{2}{qM\alpha^2}P(x)\,x^2~,
\end{equation}
where $P$ is the net effective power; now it also receives a contribution  
\begin{equation}\label{P_DC}
P_{DC}=\frac{q\alpha^2}{2(1+q)x}\frac{d\tilde M_c}{dt}~,
\end{equation}
from the mass change of the system due to the cloud's GW emission, which can be approximated as an isotropic mass loss of the host BH; see Appendix~\ref{Appendix_C}. Hence in the present case,
\begin{equation}
P_{DC}=-\frac{\tilde M_c}{M_c}\frac{q\alpha^2}{2(1+q)x}P_{gw,c}~,
\end{equation}
and
\begin{equation}
P=P_{gw}+P_{others}+P_{DC}~.
\end{equation}
For $P_{others}$, we examine the following effects.

\subsubsection{Mode-Mixing and Dynamical Friction}
In the presence of a companion body, there can be ``global'' exchange of angular momentum between the cloud and the orbit mediated by the companion's gravitational potential (including the potential of an inertial acceleration due to the orbit motion) $V_*(t,\mathbf{r})$. In the Newtonian regime, the companion's gravitational influence is fully captured by adding $V_*$ to the Schrodinger equation of the bosonic field \cite{B1,Tomaselli:2023ysb}. The resulted cloud evolution due to nonzero level mixing $H_{ab}=\langle\boldsymbol{\Psi}_a|V_*|\boldsymbol{\Psi}_b\rangle$ backreacts on the orbit dynamics and leads to rich phenomenology. There are two types of mixing: the mixing between bound states, and the mixing between a bound state with continuum states. The former is responsible for the resonance effects occurring at a discrete set of orbit frequencies $\Omega=\frac{\Delta \omega_R}{\Delta m}$ \cite{B1,B2,B3,Berti_1,HSC} and the modifications of the cloud's superradiant instabilities from nonresonant mode mixings \cite{WY_3}, while the latter leads to a continuous orbit dissipation from the ``ionization" of the bound state \cite{B4,B5,ZJ,Takahashi_2,Tomaselli:2023ysb}.

When the binary orbit frequency is off resonance, the effect from bound-state mixing is expected to be unimportant at least for the SR ground state \cite{ZJ}. In \cite{B4,Tomaselli:2023ysb} it has been argued that ionization is actually the manifestation of dynamical friction in the GA system. Since the ionization for higher-spin field has not yet been calculated, in this work we would still use the model of \cite{Hui:2016ltb} to estimate the consequence of DF. In this DF model, a test body traveling in a nonrelativistic scalar field background with relative velocity $V$ experiences a gravitational drag force,\footnote{Note that this formula was originally derived for an infinite and homogeneous background, here we apply it to the local scalar field of GA as an approximation. An extension of this result for BH moving at relativistic speeds can be found in \cite{Traykova:2023qyv,Vicente:2022ivh}. The effects of dynamical friction on the binary in ultralight scalar field environment have also been studied in \cite{Annulli:2020ilw,Annulli:2020lyc}.}
\begin{equation}\label{DF_model}
	\mathbf{F}_{DF}=-\frac{4\pi M_*^2\rho(\mathbf{r})}{V^3}C_\Lambda(\xi,\mu Vr_\Lambda)\,\mathbf{V}~.
\end{equation}
Treating scalar GA as the environment, $\mathbf{V}$ is the companion's velocity relative to the cloud, $V=\left|v \mp \frac{m}{\mu r}\right|=
\alpha\left| \sqrt{(1+q)x}  \mp m\right|x^{-1}$, where the plus (minus) sign corresponds to counterrotating (corotating) orbit. For large radius $V$ is dominated by the orbit velocity $v$, and $\xi\equiv \frac{M_*\mu}{V} \approx \frac{q\sqrt{x}}{\sqrt{1+q}}$. The uncertainly of DF lies mainly in the estimation of $C_\Lambda$, which in the present problem depends solely on the orbit radius (for circular orbit the DF force is also expected to have a radial component \cite{Buehler:2022tmr}, it is however irrelevant to the orbit dissipation). In this model it is given by \cite{Hui:2016ltb}
\begin{equation}
C_{\Lambda}(y)=\operatorname{Cin}(2 y)+\frac{\sin 2 y}{2 y}-1~,
\end{equation}
for $\xi\ll 1$, here $\operatorname{Cin}(z)=\int_0^z(1-\cos t)dt/t$. Following \cite{ZJ} we choose the IR regularization scale $r_\Lambda$ to be the cloud size measured by $r_{97}=x_{97}r_c$ (the radius encompassing 97\% of the cloud mass) for orbit radius $x>x_{97}$, hence $y\equiv\mu vr_\Lambda=\sqrt{\frac{1+q}{x}}x_{97}$. The resulted dissipation power due to DF is then
\begin{equation}\label{p_df}
	P_{DF}=-\mathbf{F}_{DF}\cdot\mathbf{v}
	=\frac{4\pi q^2M^2}{\alpha\sqrt{1+q}}\rho(x)C_\Lambda\sqrt{x}\equiv \frac{q^2\alpha^5\beta}{\sqrt{1+q}}\mathcal{P}(x)~.
\end{equation}
Since the nonrelativistic Coulomb scattering problem (based on which the DF above is derived \cite{Hui:2016ltb}) is same for each component of the wave function of a higher spin field, the result can be generalized simply with $\rho=M_c\text{Tr}(\boldsymbol{\Psi}^\dagger\boldsymbol{\Psi})$.

\begin{figure}[t]
	\centering
	\includegraphics[width=0.5\textwidth]{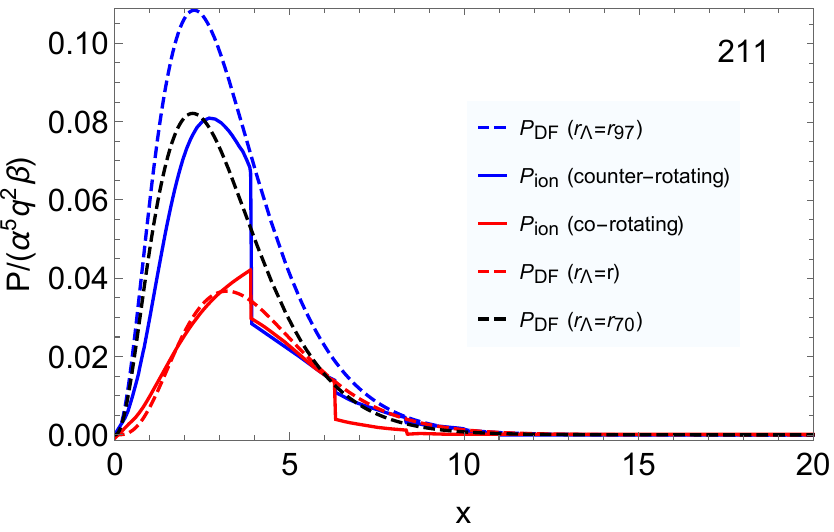}\quad
	\includegraphics[width=0.5\textwidth]{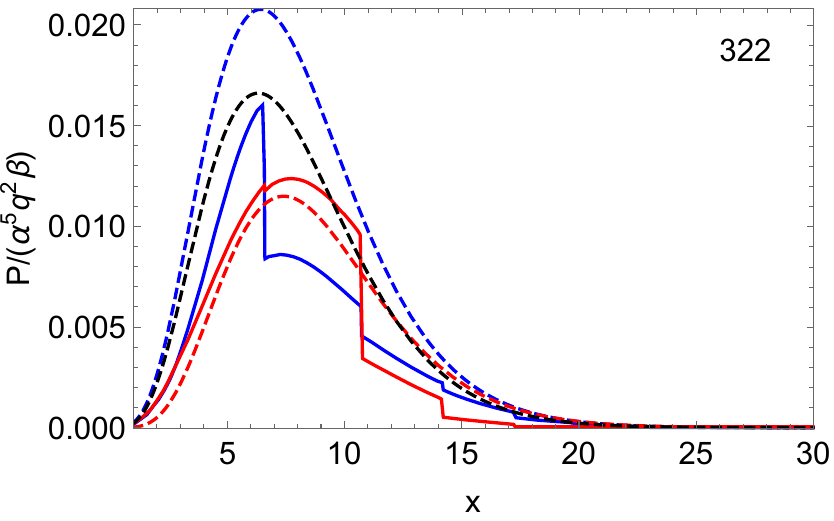}
	\caption{Comparison of DF power Eq.~\eqref{p_df} with ionization power of scalar 211 and 322 states for equatorial plane circular orbit.}\label{DF}
\end{figure}

Strikingly, we find that for scalar cloud this estimation (with $V\approx v$) agrees well with the ionization power calculated in \cite{Tomaselli:2023ysb} in the overall trend and magnitude; see Fig.~\ref{DF}. Indeed, the scaling form of ionization power is the same with Eq.~\eqref{p_df} for $V\approx v$ if $q\ll 1$ (the result only changes slightly at the small radius after including the cloud velocity in DF model). This demonstrates that the DF and the ionization model are indeed compatible, though the DF model tends to overestimate the orbit dissipation [especially for the corotating orbit; a smaller value of $r_\Lambda$ fits better with $P_{ion}$ in Fig.~\ref{DF}, but the relative difference is within $\mathcal{O}(1)$] and without the features of discontinuity\footnote{Such discontinuity originates from the nonzero mode-mixing between the given bound state being ionized and a continuum state with zero wave-number \cite{B4}.}, also in the ionization model the cloud is being consumed. For higher-spin field, the ionization power has not yet been calculated, but since the difference lies mainly in the angular mixing, we expect a similar result.

\subsubsection{Accretion}\label{accretion_compare}
If the companion is a BH, besides friction, additional drag force arises due to its accretion of the ambient cloud. In a uniform background of ultralight scalar field, the force due to accretion is $\mathbf{F}_{acc}=-\dot M_*\mathbf{V}$ with $\dot M_*\equiv \sigma \rho V$. For $V>2\pi M_*\mu$ the absorption cross section can be approximated \cite{Benone:2019all} as $\sigma=A/V$, where $A\sim 16 \pi M_*^2$ is the area of BH horizon (see also \cite{B4}), while for $V<2\pi M_*\mu$ the result is $\sigma=\frac{32\pi^2M_*^3\mu}{V^2}$. The effective powers due to accretion in the two regimes are 
\begin{equation}\label{acc1}
P_{acc}=16\pi q^2M^2\alpha^2\rho(x)\, x^{-1}~,
\end{equation}
and
\begin{equation}\label{acc2}
P_{acc}=32\pi^2q^3M^2\alpha^2 \rho(x)\, x^{-1/2}~.
\end{equation}
The accretion powers from both estimations are strongly suppressed relative to the dynamical friction as
\begin{equation}
\frac{P_{acc}}{P_{DF}}<\frac{\alpha^3}{C_\Lambda}~.
\end{equation}
and the effect from cloud mass loss due to accretion is even smaller, suppressed relative to $P_{acc}$ by $q^2$. Although there are currently no quantitative computations for the accretion rate of higher-spin massive bosonic fields, we expect the result will be at the same order of magnitude. Hence we shall neglect the companion's mass accretion in the following.

\section{Binary Evolution}\label{sec_3}
In this section we analyze the binary orbit evolution under DC and DF. The GA is approximately rigid provided that it is off resonance and the companion's perturbation is small, $|V_*/\left(-\frac{\alpha}{r}\right)|\lesssim q\left(\frac{x_{97}}{x}\right)^3\ll 1$. Therefore, besides the extreme-mass-ratio system with $q\ll 1$, this model can also be applied to binary with larger mass ratio so long as $x$ is sufficiently large \cite{Takahashi_1}. Since the innermost stable circular orbit (ISCO) radius $x_{isco}=6\alpha^2$ of the host BH is deep inside the cloud where the binary may subject to strong mode-mixing or even nonperturbative effects, our discussion would be restricted to the phase of orbit evolution at large radius; specifically we would take $x>10$.

The evolution of circular orbit is given by Eq.~\eqref{orbit}, where the power function $P(x)$ depends solely on $\beta, \alpha$ and $q$, so the BH mass $M$ only affects the overall timescale\footnote{This is the case even if the time-dependent depleted value of $\beta$ is used, since the depletion timescale is also proportional to $M$.}. At radius $x$, the binary GW frequency is
\begin{equation}\label{freq}
\begin{aligned}
f&=\Omega/\pi=\left[\frac{(1+q)^{1/2}\alpha^3}{\pi M}\right]x^{-3/2}\equiv \kappa x^{-3/2}
\\
&=(1+q)^{1/2}\frac{M_\odot}{M}\left(\frac{\alpha}{0.1}\right)^3\left(\frac{10}{x}\right)^{3/2}\times 2\,\text{Hz}.
\end{aligned}
\end{equation}
From Eq.~\eqref{orbit}, we have
\begin{equation}
	\dot{f}=M^{-5/3}\left[\frac{3(1+q)^{1 / 3}}{\pi^{2 / 3} q}\right] P(x) f^{1 / 3}~.
\end{equation}
The deviation of $P$ from $P_{gw}$ could then be observed in the binary GW signal or through high-precision pulsar timing, if the companion happens to be a pulsar (PSR). A characteristic measure for the frequency change is the braking index, which in the case of circular orbit can be written directly with the effective power:
\begin{equation}
n_b\equiv\frac{\dot{\Omega} \ddot{\Omega}}{\dot{\Omega}^2}=\frac{5}{3}-\frac{2}{3} \frac{x \ddot{x}}{\dot{x}^2}=\frac{1}{3}-\frac{2}{3}\frac{x P'}{P}~.
\end{equation}
For $P=P_{gw}$, $n_b=\frac{11}{3}$, while for $P=P_{DC}$ and assuming an approximately constant mass depletion rate, $n_b=1$. Another useful measure is the overall GW dephasing
\begin{equation}
\Delta \Phi (t)=\Phi(t)-\Phi_{GR}(t)=2\pi \int_0^t dt'\,\left[f(t')-f_{GR}(t')\right]~,
\end{equation}
where $[0,t]$ is the time span of an observation and $f_{GR}(t)$ is given by the vacuum evolution.

\subsection{Early Inspiral}
\begin{figure*}[ht]
	\centering
	\includegraphics[width=0.47\textwidth]{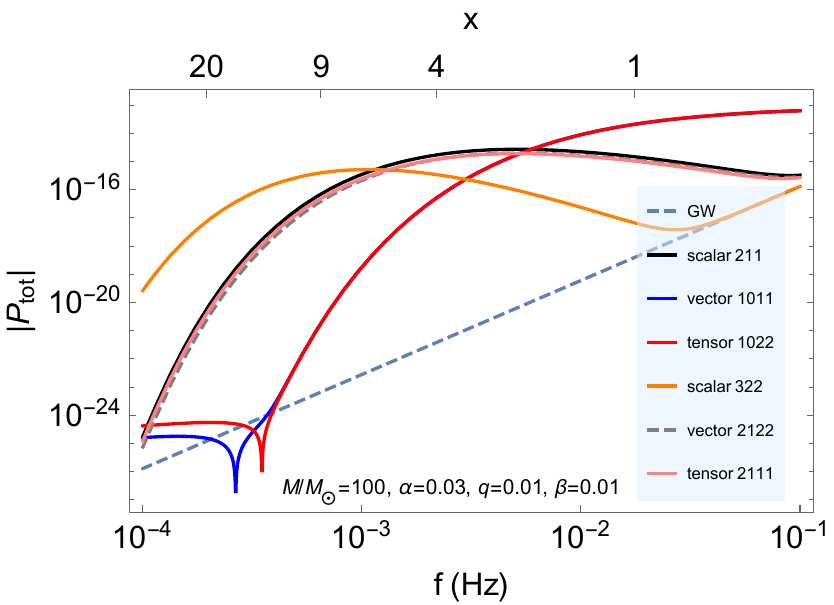}
	\quad
	\includegraphics[width=0.46\textwidth]{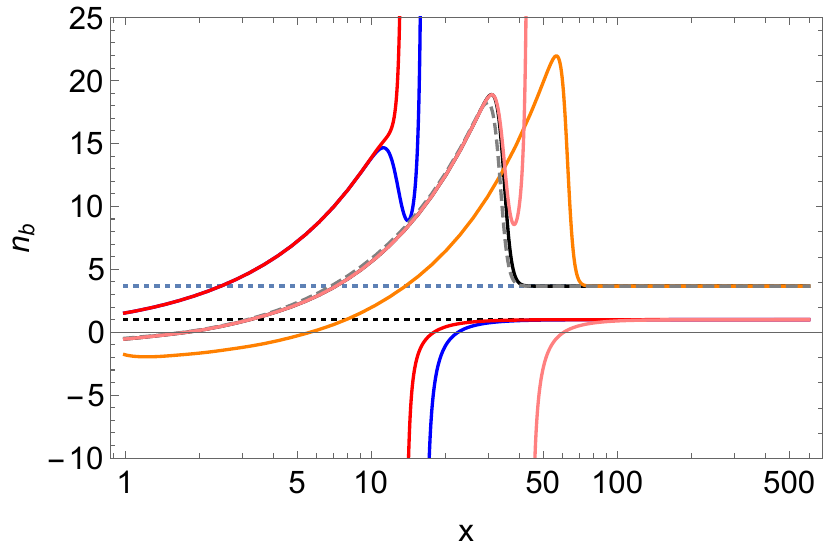}
	\quad
	\includegraphics[width=0.47\textwidth]{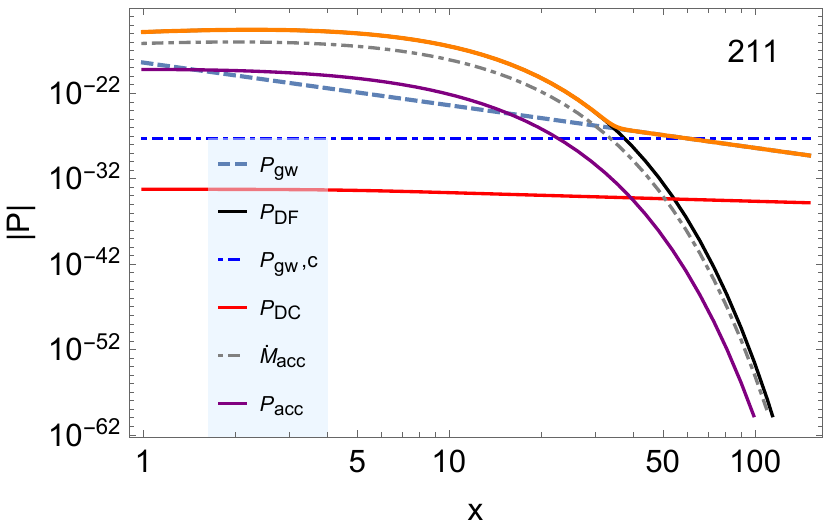}
	\quad
	\includegraphics[width=0.47\textwidth]{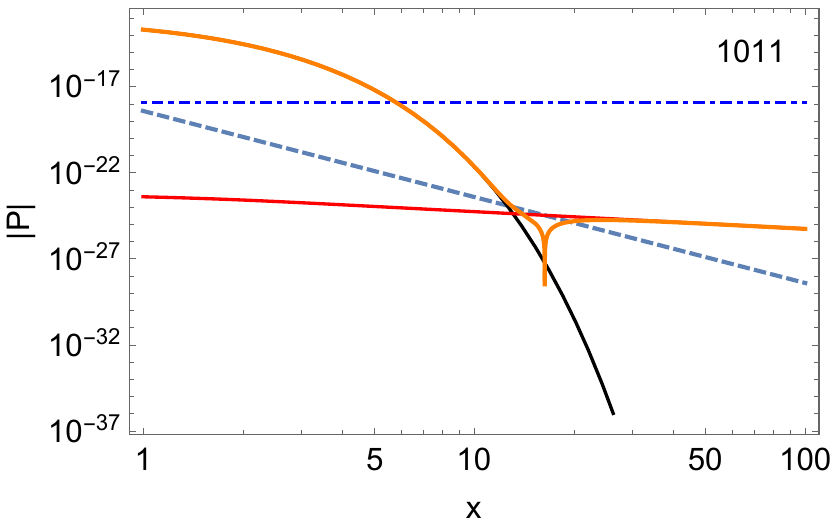}
	\caption{Upper left: the effective power $|P(x)|$ for GA saturated in various states; the sharp dip is due to the sign change and represents the critical radius of outspiral, and the dashed line is the vacuum power $P_{gw}$. Upper right: the corresponding braking index, for $P=P_{gw}$, $n_b=\frac{11}{3}$, while for $P=P_{DC}$ with approximately constant cloud mass, $n_b=1$, as shown by the dotted lines. Lower plots depict the components of the net power for scalar 211 and vector 1011 state, where we also show the effective power due to accretion according to the estimation Eq.~\eqref{acc1} for the scalar GA. It is seen that accretion into the companion could lead to much stronger cloud mass loss (gray dot-dashed line) at small orbit radius comparing to the intrinsic mass depletion of the cloud (blue dot-dashed line), where however the orbit evolution is dominated by the DF. For this set of parameters, $\mu=4\times 10^{-14}$eV (the frequency of the GW emitted from the cloud is $\mu/\pi=19$ Hz), and $x_\text{ISCO}=0.0054$.}\label{power_example}
\end{figure*}

For the companion to inspiral, the combined dissipation power from binary GW emission and the DF should overwhelm the negative power of cloud mass depletion, which is the case for a sufficiently small orbit radius or cloud mass. Typically the DF power could strongly dominate over $P_{GW}$ for small radius with $x>1$, which might lead to a considerable amount of GW dephasing and a shorter merger time. The situation at larger radius depends on the state and the cloud mass; e.g., among the six states depicted in Fig.~\ref{power_example}, for scalar 211, 322 and vector 2122 states there is an intermediate $P_{GW}$ domination spanning a broad range of radius, followed by a transition to $P_{DC}$ domination at even larger radius. While for the other states of vector and tensor atom the region of $P_{GW}$ domination is negligible (for smaller cloud mass, it is broadened). As shown by the power ratios for $q\ll 1$:
\begin{equation}
\frac{P_{DF}}{P_{gw}}\sim \frac{\beta}{\alpha^5}x^5\mathcal{P}(x)~,\quad
\frac{P_{DC}}{P_{gw}}\sim \frac{\beta ^2p(\alpha)}{\alpha^8q}x^4~,
\end{equation}
$P_{DC}$ is enhanced for smaller $q$, but suppressed for smaller $\alpha$ and $\beta$ relative to $P_{DF}$.

\subsection{Outspiral}

Since $P_{DC}$ always dominates at sufficiently large radius, there would be a critical radius beyond which the companion outspirals. The critical radii around GA in various SR states computed with the full power model are shown in Fig.~\ref{x_crit}. We notice that for scalar 211, 322 and vector 2122 state, the critical radius is completely fixed by the balance between $P_{gw}$ and $P_{DC}$ such that
\begin{equation}\label{x_crit}
x_{crit}=\left[\frac{64}{5}\frac{q(1+q)^2\alpha^8}{p(\alpha)\beta^2}\right]^{1/4}~.
\end{equation}
For states with large $p(\alpha)$ (vector 1011, tensor 1022 and tensor 2111), the critical radius is enlarged as compared to the result without DF since Eq.~\eqref{x_crit} dives into smaller radius where orbit dissipation is stronger. By the same reason, this enlargement is stronger for smaller mass ratio $q$ and larger cloud mass $\beta$. For small enough $\beta$, the critical radius is still given by Eq.~\eqref{x_crit}.

The transition to $P_{DC}$ domination turns out to be rather sharp. For $x>x_{crit}$ (or for $x<x_{crit}$ before $P_{DF}$ becomes important), the power is well approximated by $P_{gw}+P_{DC}$; the resulted orbit evolution is
\begin{align}\label{eq_full}
\dot x &= -Ax^{-3}+\frac{B}{(1+\frac{t}{\tau_{gw}})^2}x~, \\
  A&\equiv \frac{64}{5}\frac{\alpha^{8} (1+q)q}{M},\quad B\equiv \frac{\beta^2 p(\alpha)}{M(1 + q)}.\nonumber
\end{align}
where $\beta$ is the cloud mass ratio at some initial time. With $x(0)=x_0$, this equation admits an analytical solution :
\begin{align}\label{x_full}
    x^4(t)=&e^{-\frac{4 B \tau_{gw}^2}{t+\tau_{gw}}} \left\{16 A B \tau_{gw}^2 \left[\text{Ei}\left(\frac{4 B \tau_{gw}^2}{t+\tau_{gw}}\right)-\text{Ei}(4 B \tau_{gw})\right]\right. \nonumber \\
    &+e^{4 B \tau_{gw}} \left(4 A \tau_{gw}+x_0^4\right)\Bigg{\}}-4 A (t+\tau_{gw})~,
\end{align}
describing both inspiral and outspiral, where $\text{Ei}(x)=-\int_{-x}^\infty dz\frac{e^{-z}}{z}$ is the exponential integral. In timescale much shorter than $\tau_{gw}$, $\beta$ is approximately constant and this simplifies to
\begin{equation}\label{x_outspiral}
	x(t)=\left[\left(x_0^4-\frac{A}{B}\right)e^{4Bt}+\frac{A}{B}\right]^{1/4}~.
\end{equation}
The corresponding GW phase is given by
\begin{equation}
		\Phi(t)=2\pi\int_0^t dt'f(t')=
		\left.
		-\frac{4\pi \kappa}{3B}\frac{_2F_1\left(\frac{3}{8},\frac{3}{8};\frac{11}{8};\frac{-A e^{-4 B t'}}{B x_0^4-A}\right)}{e^{3Bt'/2}(x_0^4-A/B)^{3/8}}
		\right|^t_0~,
\end{equation}
where $\kappa\equiv \frac{(1+q)^{1/2}\alpha^3}{\pi M}$. If the orbit evolution is purely driven by cloud depletion, this simplifies to
\begin{equation}
	\Phi(t)=
	-\frac{4\pi \kappa}{3 B x_0^{3/2}}(e^{-3 B t/2}-1)~.
\end{equation}
While for ordinary binary inspiral with $B=0$, the phase is $\Phi_{GR}(t)=4\pi \kappa \left[ \left(4 A t+x_0^4\right)^{5/8}-x_0^{5/2} \right]/(5A)$.
\begin{figure*}[t]
	\centering
	\includegraphics[width=0.3\textwidth, height = 0.25\textwidth]{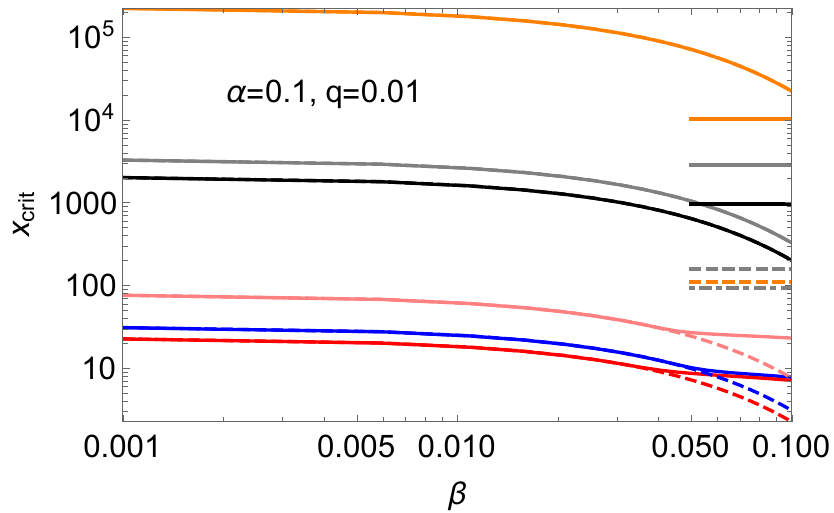}
	\includegraphics[width=0.3\textwidth, height = 0.25\textwidth]{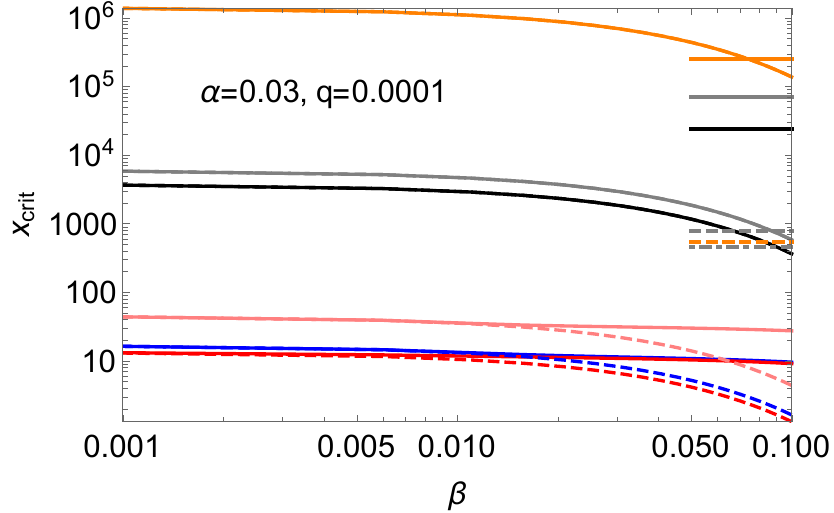}
	\includegraphics[width=0.38\textwidth, height = 0.252\textwidth]{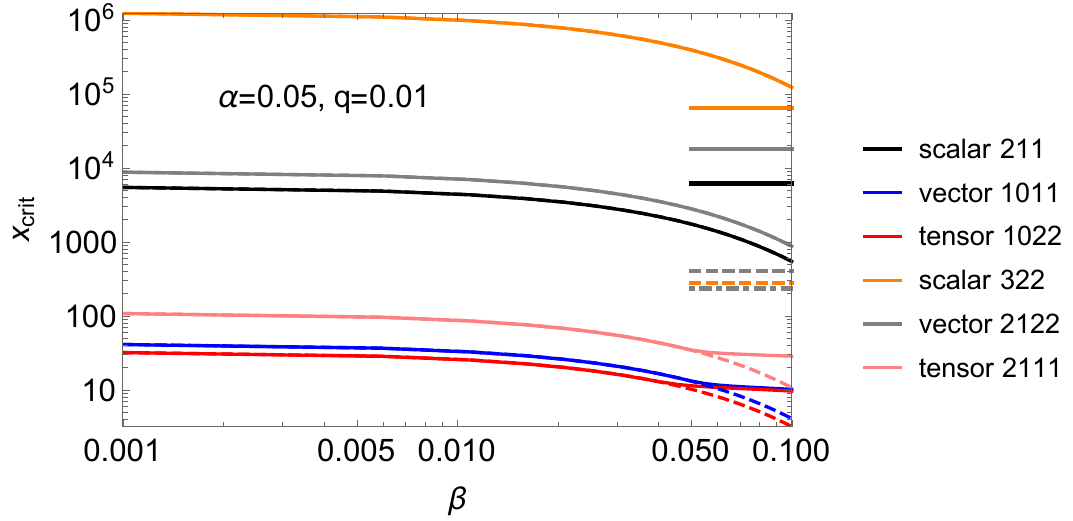}
	\caption{ Critical radius of outspiral, in the presence (solid lines) or absence (dashed lines) of dynamical friction. The short horizontal lines are the radii of fine and hyperfine resonances (for saturated BH spin $a=4\alpha/m$ in the $m$ state): 211 to 21-1 (black), 322 to 300 (orange dashed) and to 320 (orange solid), 2122 to 2100 (gray dot-dashed), 2110 (gray dashed) and 2120 (gray solid).}\label{x_crit}
\end{figure*}

\paragraph{Secular Evolution} 
Once the companion outspirals, it will not return until a sufficient depletion of the cloud. To track the long-term orbit evolution, we must inspect on the full solution Eq.~\eqref{x_full}. For outspiral, since $x(t)<x_0e^{Bt}$, the cumulated orbit radius change $\Delta x<x_0 (e^{Bt}-1)$. Since $B=\frac{\beta}{\tau_{gw}(1+q)}$, this means roughly that the fractional orbit radius change after time $\tau_{gw}$ cannot be larger than $\beta$. We find this is actually a good estimation for the \textit{maximum}
value of $(x-x_0)/x_0$ attainable during the outspiral. For large orbit radius Eq.~\eqref{x_full} is well approximated by
\begin{equation}\label{x_approx}
x(t)\approx 
\left[
-4A(t+\tau_{gw})+e^{\frac{4B \tau_{gw}}{1+\tau_{gw}/t}}(4A\tau_{gw}+x_0^4)
\right]^{1/4},
\end{equation}
i.e., without the exponential integrals, and for sufficiently large $x_0$ this indeed approaches to $x_0(1+\beta)$. Also from this approximation we can find a good estimation for the maximum time of outspiral,
\begin{equation}\label{t_approx}
t_{max}\approx \left[-1-2B\tau_{gw}+\sqrt{B(4\tau_{gw}+x_0^4/A)}\right]\tau_{gw}.
\end{equation}
At large radius the time of outspiral can be much longer than $\tau_{gw}$, though it would still be a ``transient'' phase comparing with the subsequent inspiral. 

\begin{figure}[hbt]
	\centering
	\includegraphics[width=0.47\textwidth]{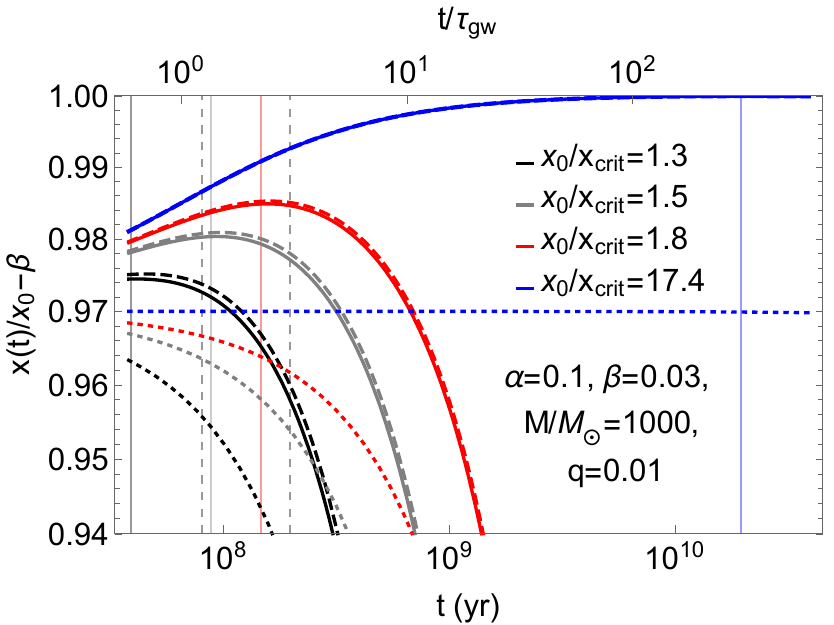}
	\caption{Orbit evolution during outspiral around GA in the scalar 211 state for various initial radii. The solid line corresponds to the full solution Eq.~\eqref{x_full}, the dashed line for the approximated solution Eq.~\eqref{x_approx}, and the dotted line for the vacuum solution without mass depletion. The approximation Eq.~\eqref{t_approx} for the maximum time of outspiral is shown in solid vertical line. The two dashed vertical lines are $t=\Delta t +\tau_I^{(322)}$ and $\Delta t +2\tau_I^{(322)}$, respectively. Here $\tau_I^{(322)}$ is the instability timescale of 322 state under saturated BH spin of 211 state, and $\Delta t=(1-\alpha/\beta)\tau_{gw}$ is approximately the time interval between the initial time of outspiral and the onset of 211 saturation.}\label{211_out}
\end{figure}

As a concrete example, we consider the orbit evolution around GA in scalar 211 state; the results are shown in Fig.~\ref{211_out}. It is seen that for large radius the timescale of outspiral can even be comparable with the timescale of the growth of 322 state, which might bring the outspiral to an early end since the 322 state decays much slower than the 211 state, though an accurate description of such processes awaits for more detailed investigations.

\paragraph{Constraint on Resonances} Since the binary orbit during outspiral undergoes little frequency change, this also implies that resonance event is unlikely to take place during the outspiral. For example, the resonance between scalar 211 and 21-1 states, a hyperfine transition, is at radius
\begin{equation}
x_*=\left[\frac{144(1+q)}{\chi^2}\right]^{1/3}\alpha^{-2}.
\end{equation}
For this transition to happen we require at least that $x_*<x_{crit}$, which translates into a maximum cloud mass before the resonance
\begin{equation}
\beta_{max}=0.13\,\chi^{4/3}(1+q)^{1/3}q^{1/2}\alpha^8p^{-1/2}.
\end{equation}
For $q=10^{-3}$ and $\chi=4\alpha$, $\beta_{max}=1.3\times 10^{-3}$ if $\alpha=0.1$ while $\beta_{max}=5\times 10^{-6}$ if $\alpha=0.01$.

Similar constraints can be put on the other possible transitions. The leading quadrupole transitions for the scalar 322 state are the Bohr transitions ($n'\ne n$) to 200 and 100, fine transition ($n'=n$, $l'\ne l$) to 300, and the hyperfine transition ($n'=n,\,l'=l,\,j'=j$, $m'\ne m$) to 320. For vector 1011 state there are only Bohr transitions to 321-1,322-1,323-1 and 3233 (and higher $n$ states), similarly for tensor 1022 state to 3200, 3210, 3220, 3230, 3240 and 3244; for vector 2122 state, the Bohr transitions to e.g., 3100, 4320, 3110, 4330, 3120, 4340 and 4344, and (hyper)fine transitions to 2100, 2110 and 2120 are possible. As pointed out in \cite{WY_2}, the resonant orbit frequency for Bohr transition $\sim \mathcal{O}(\mu\alpha^2)$, corresponding to the orbit radius $x\sim\mathcal{O}((1+q)^{1/3})$, which might invalidate the perturbative model of GA, so we consider only the fine and hyperfine transitions at larger radius, which are depicted in Fig.~\ref{x_crit}. We find that $\beta_{max}$ decreases with decreasing mass ratio $q$ and for hyperfine transitions also with decreasing $\alpha$.

In this discussion we have neglected the companion-induced corrections to $\omega_I$ of the SR state. As noted by \cite{WY_3} such corrections are negative. Hence for a given BH spin there is a critical radius below which the effective value of $\omega_I$ turns negative and the cloud gets reabsorbed from mixings with decaying modes. Actually for a saturated state by definition $\omega_I= 0$, the cloud reabsorption is unavoidable, the strength of such effect is proportional to the perturbing potential $V_*$ so it is less important at large radius, especially for SR ground state which does not mix with the relatively fast-decaying lower-$l$ states in the leading quadrupole order.

\section{Observability}\label{sec_4}
We proceed to assess the detectability of the cloud-induced DC and DF effects (namely $P-P_{gw}$) on the orbit phase evolution. We focus on the direct observation of outspiral and the detecting threshold in the case of BH-PSR binary. From Eq.~\eqref{freq} it can be seen that for a BH with $M>10^4M_\odot$, the signal for radius $x>10$ and $\alpha\sim \mathcal{O}(0.1)$ is typically lower than the observation window $(10^{-4},10^{-1})$~Hz of the space-borne gravitational-wave interferometers. If the host BH of the cloud is lighter, however, the effect of DF and DC may be observed directly in the sensitive band of GW detectors, once such an event is detected.

For a very small mass ratio, $q\ll 1$, such as an intermediate mass ratio ($10^{-2}<q<10^{-4}$) or extreme mass ratio ($q<10^{-4}$) binary system, $P_{DF}/P_{gw}$ is insensitive to $q$, while $P_{DC}$ is enhanced by $q^{-1}$. For mass ratio as large as $q\sim 1$, our model is valid only for $x \gg 1$, where DF is expected to be unimportant. In an ordinary binary, the companion can be a stellar mass object with $M_*=0.1\sim 100M_\odot$. But we can also consider the possibility of the companion being a very light primordial black hole (PBH). Note that if the mass of the PBH is too small, its own mass loss through Hawking evaporation may also need to be taken into account \cite{Blachier:2023ygh}.

There are no fundamental restrictions on the values of $\alpha$ and $\beta$ except that $\beta$ should be smaller than its initial saturated value, but $\alpha$ cannot be too small if we require a finite formation time of the cloud perhaps within the binary lifetime, for that we shall impose $\tau_I<10^6\text{yr}$. On the other hand, the maximum value of $\alpha$ would be limited if we require a sufficiently long cloud depletion time. $\tau_{gw}$ should be much larger than $\tau_I$, moreover, in \cite{B3} the bound $\tau_{gw}(M_c=M_{c,0})>10^8\text{yr}$ was adopted to guarantee the stability of the cloud in astrophysical timescale. But the mass depletion could continue within a time much longer than $\tau_{gw}$, just leading to smaller existing cloud mass. In the following we consider a relaxed bound with $\tau_{gw}(M_c=M\beta)>10^4\,\text{yr}$, demonstrating how the detectable parameter space is squeezed by the requirement of a minimum depletion timescale.

\subsection{Observing the Outspiral}
\begin{figure*}[hbt]
	\centering
	\includegraphics[width=0.47\textwidth]{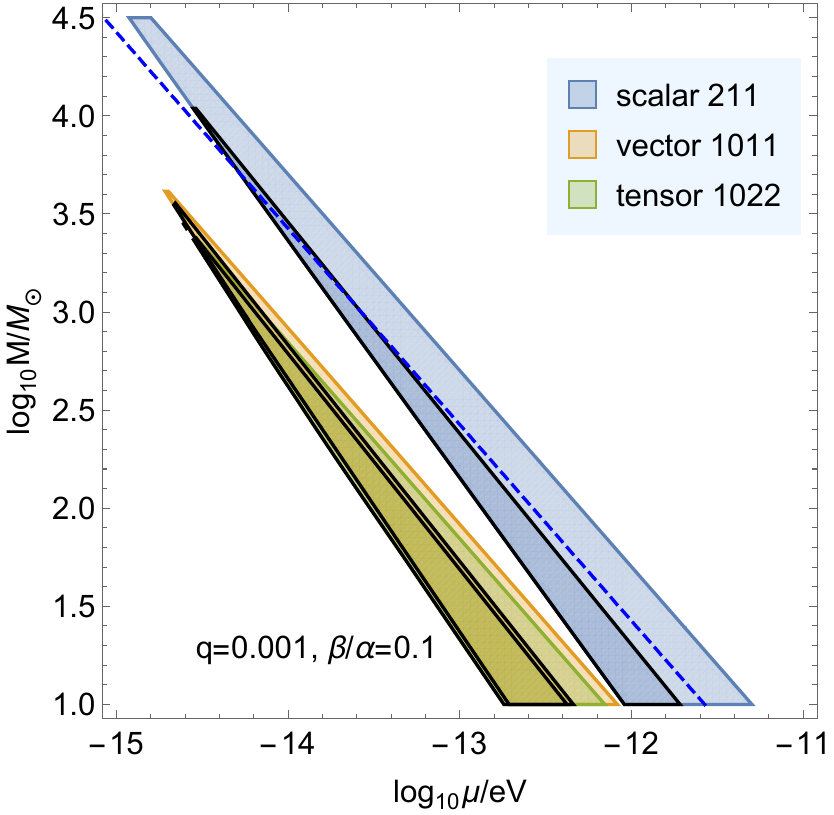}
	\quad
	\includegraphics[width=0.47\textwidth]{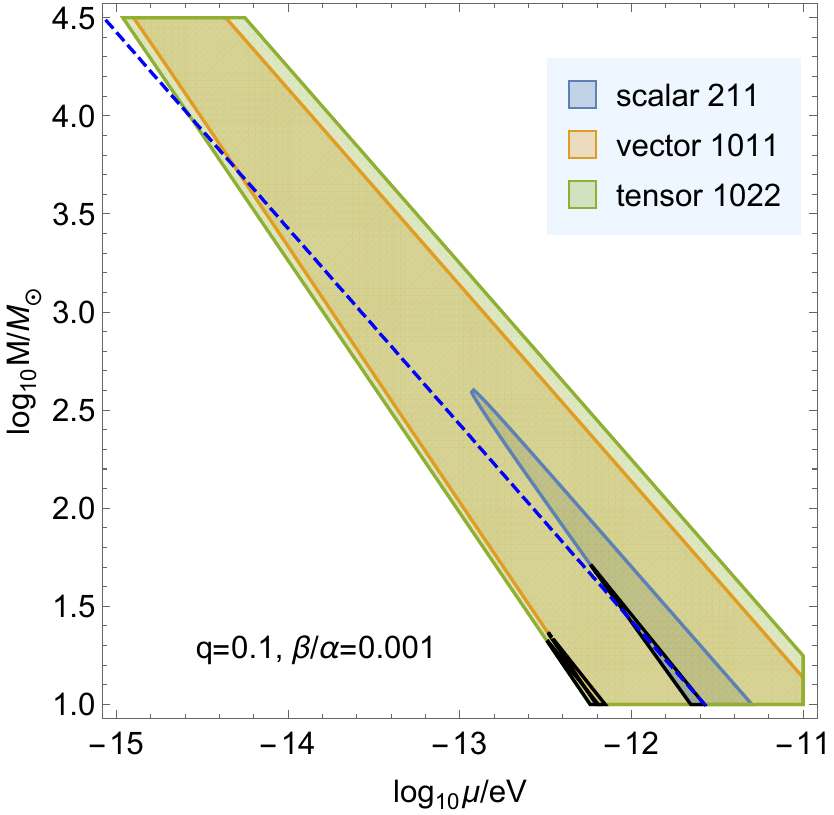}
	\caption{Region of BH mass and bosonic particle mass satisfying $f(x_{crit})>10^{-4}$Hz and $x_{crit}>10$ for given binary mass ratio $q$ and $\beta/\alpha$. The regions with black boundary are the results after imposing the constraints $\tau_{gw}>10^{4}$ yr and $\tau_{I}<10^6$ yr. The blue dashed line corresponds to $\alpha=0.2$.}\label{out-spiral}
\end{figure*}

By requiring that $f(x_{crit})>10^{-4}\,\text{Hz}$ we can estimate the mass range of the bosonic particle for which the outspiral can possibly be observed by the space-borne GW detectors such as LISA and Taiji. The results are presented in Fig.~\ref{out-spiral}, where we can see that the mass range of $\mu$ for a given BH mass shrinks for larger $M$, the lower bound corresponds to $f(x_{crit})=10^{-4}\,\text{Hz}$ and the upper bound comes from the constraints on $x_{crit}$ and $\tau_{gw}$, the latter being more stringent. The parameter space of vector 1011 and tensor 1022 states largely coincide, since they have a similar depletion rate. As shown in the last section, the critical radius of outspiral is typically given by Eq.~\eqref{x_crit}:
\begin{equation}
x_{crit}^4\propto \alpha^8 \tau_{gw} ^2 p(\alpha)=M^2\beta^{-2}p^{-1}\alpha^8.
\end{equation}
Assuming a fixed value for the BH mass, for given $\tau_{gw}$ and $\alpha$, the critical radius increases with $p$. But for given $x_{crit}$ and $\beta$, $\alpha$ decreases with $p$. Then since the depletion rate of vector and tensor SR ground state for same $\alpha$ is larger than that of scalar, the minimum mass of vector or tensor boson supporting outspiral around the same host BH at given orbit frequency is lighter. On the other hand, for fixed $\tau_{gw}$ and $\beta$, the mass of vector or tensor boson is also lighter than the scalar boson.

\subsection{Pulsar Timing Detection}
If the companion is a pulsar, pulsar timing provides an accurate way to measure the orbit evolution \cite{WY_1,WY_2,WY_3,Akil:2023kym}. For definiteness, we choose the benchmark values $M_*=1.6M_\odot$. The detection threshold is \cite{WY_1}
\begin{equation}
|\Phi(t)-\Phi_{GR}(t)| > 4\pi \sigma(t),
\end{equation}
with the uncertainty of phase measurement approximately given by
\begin{equation}
\sigma(t)=\frac{1}{\sqrt{\lceil t / 1\,\text {day}\rceil}} \frac{T_{p}}{\min \left(t_{\mathrm{obs}}, t\right)},
\end{equation}
where $\lceil \rceil$ is the ceiling function, $t_{\mathrm{obs}}$ the observation time per day and $T_{p}$ the pulse period of the pulsar. The time span of observation $t$ cannot exceed the duty time of the radio telescope, and it would also be shorter than the merger time if the companion undergoes an inspiral.

\begin{figure*}[hbt]
	\centering
	\includegraphics[width=0.31\textwidth]{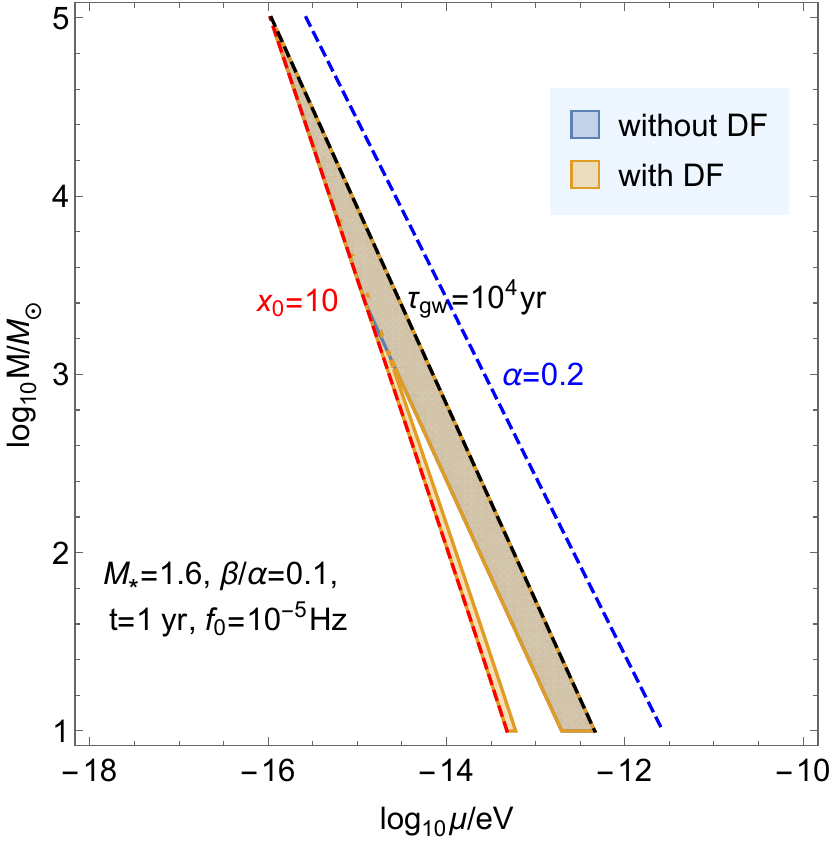}
	\quad
	\includegraphics[width=0.31\textwidth]{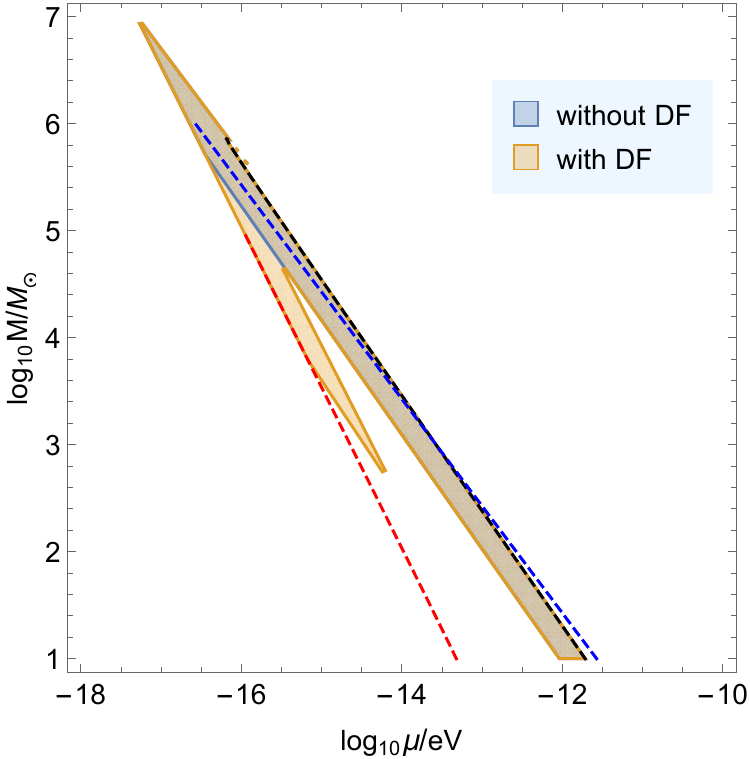}
    \quad
	\includegraphics[width=0.32\textwidth]{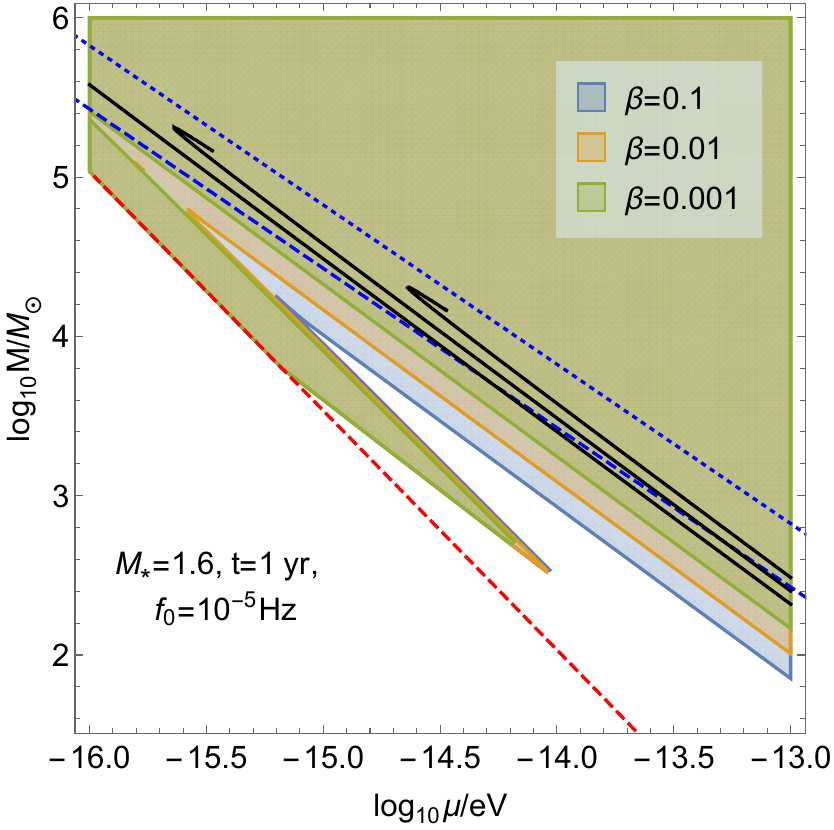}
	\caption{Region of BH mass and bosonic particle mass (left, vector 1011; middle, scalar 211) where $x_0>10$ and the deviation from vacuum orbit evolution is detectable via pulsar timing for $f_0=10^{-5}$ Hz, $t_\text{obs}=10$ h and $T_p=1$ ms, the constraints $\tau_{gw}>10^{4}$ yr and $\tau_{I}<10^6$ yr are also imposed. We have checked that throughout this parameter space, $\frac{f_0}{(\dot f)_0}\ll t$, so the quadratic approximation to the phase change is valid. The right figure is a close-up for the detectable parameter space of scalar 211 GA (with DF) with different cloud mass, where the black solid line is the contour of $\tau_{gw}=10^5$ yr (the upper one corresponds to smaller $\beta$) and the blue dashed (dotted) lines correspond to $\alpha=0.2\,(0.5)$.}\label{PT}
\end{figure*}

\begin{figure*}[hbt]
	\centering
	\includegraphics[width=0.47\textwidth]{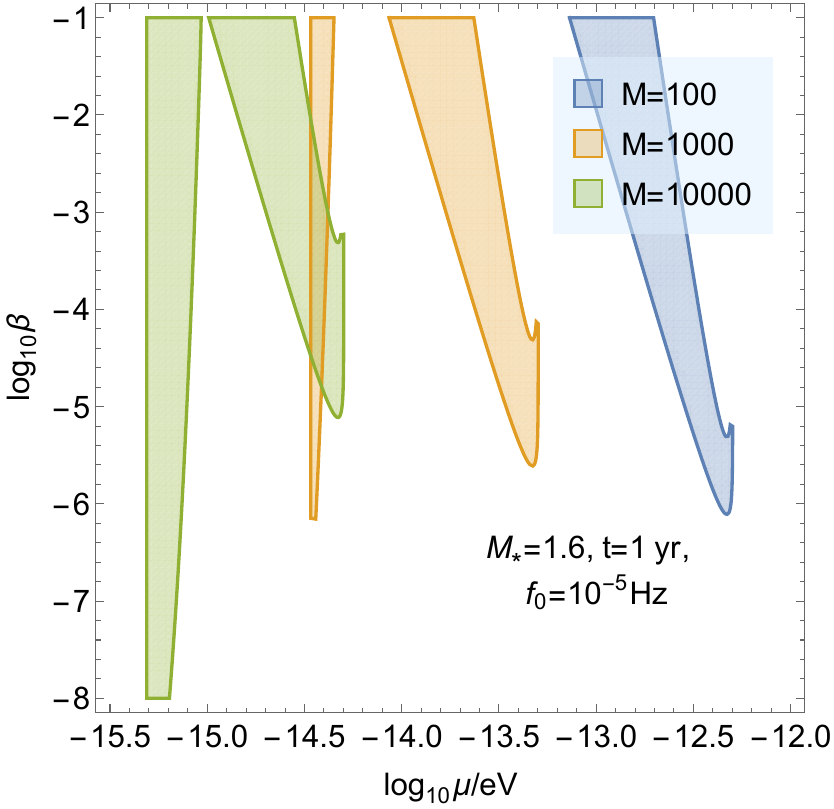}
	\quad
	\includegraphics[width=0.47\textwidth]{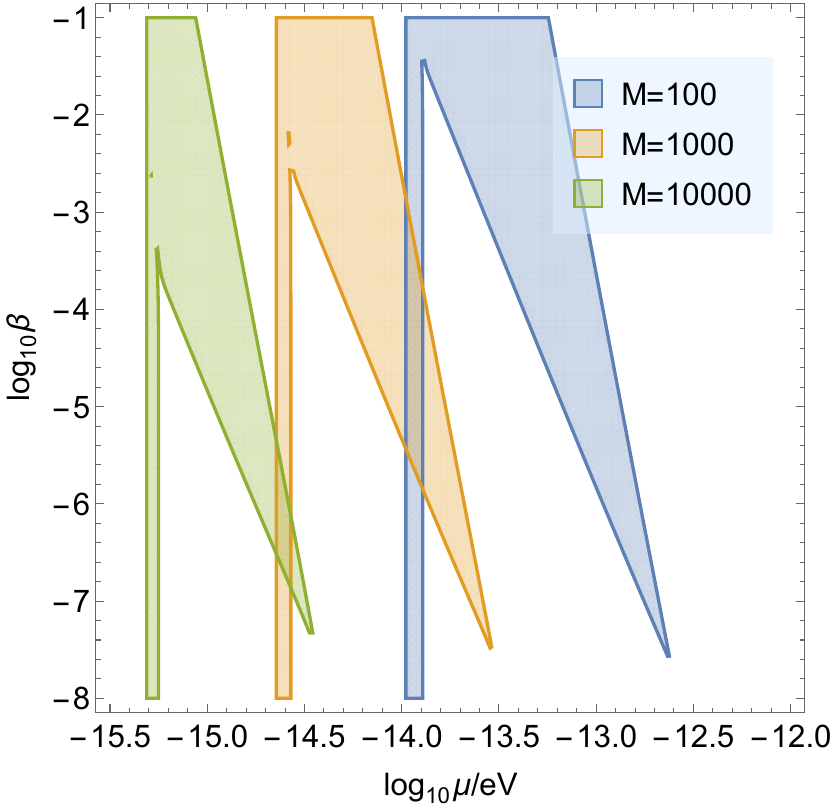}
 \caption{Detectable region of bosonic particle mass and cloud mass ratio (left, scalar 211; right, vector 1011) for different host BH mass (with DF); the constraints are the same as Fig.~\ref{PT}. Note for $\mu\sim 10^{-n}\,\text{eV}$, $\alpha\sim 10^{10-n}\times M/M_\odot$; hence for BH mass $M\sim 10^n\,M_\odot$, only the particle mass smaller than $10^{-(10+n)}\,\text{eV}$ is viable.}\label{PT2}
\end{figure*}

Assuming a large enough orbit evolution timescale, we only expect to observe the quadratic phase change $\Delta \Phi\approx 2\pi\left[f_0t+\frac{1}{2}(\dot f)_0 t^2\right]$, where
\begin{equation}
(\dot f)_0=
M^{-5/3}\left[\frac{3(1+q)^{1 / 3}}{\pi^{2 / 3} q}\right] P(x_0) f_0^{1 / 3}~.
\end{equation}
The detectable regions of $(M,\mu)$ for scalar 211 and vector 1011 states are presented in Figs.~\ref{PT} and \ref{PT2} for a fiducial set of parameters, where we show the results with or without DF; generally the upper bound comes from the constraint on $\tau_{gw}$. DF turns out to be also relevant in this low-frequency regime and signifies inspiral, since for outspiral the DF only reduces $\Delta\Phi$. The other part of the detectable region does not depend on the DF, and could be either outspiral or inspiral. For $P\approx P_{DC}+P_{gw}$, the condition of detection is explicitly given by
\begin{equation}
\frac{\beta^2p}{M(1+q)}>\frac{8\sigma}{3t^2f_0}~.
\end{equation}
Then for outspiral, again, the minimum detectable boson mass in the vector case is lighter than the scalar case for given $\beta$ and BH mass, and a more massive BH can probe lighter bosons. Also in the case, since $\sigma \sim t^{-1/2}$, the minimum detectable boson mass for a given binary is determined by $(t^{5/2}\beta^2f_0)^{-1}$, but a higher orbit frequency $f_0$ will be more constrained by the assumption $x_0>10$.

\section{Conclusions and Outlooks}\label{sec_5}
We have investigated the orbit evolution of a binary system containing a gravitational atom, based on a more comprehensive model for the binary's off-resonant orbit dynamics taking into account both the dynamical friction and the intrinsic mass depletion of the boson cloud. In this modeling, GA of spin-0, -1 and -2 bosons are treated on equal footing, which enable us to contrast the differences of the binary's effective dissipative power for the different SR states. 

One of our main motivations is to quantify the importance of the cloud mass depletion at large radius, which are relevant for early phase of binary evolution and also the resonance events. We find that DF could typically dominate at the small radius. But unless the GA is in the vector or tensor SR ground state with large enough cloud mass, the critical radius of outspiral is determined by the balance between cloud mass depletion and binary GW emission. By requiring the fine and hyperfine resonances to happen only during inspiral, upper limits are imposed on the cloud mass before the resonance, which for example can already be small for typical model parameters of the scalar 211 to 21-1 transition. We present an exact solution for the circular orbit evolution under $P_{DC}+P_{gw}$, showing that for binary system with a small mass ratio, even a small cloud mass could significantly slow down the inspiral process, or for a large enough cloud mass make the companion outspiral. This implies that not only the cloud mass depletion itself, but also the DC effects on the orbit evolution may need to be considered for a reasonable estimation of the cloud mass.

Comparing with the scalar SR state, the depletion rates of vector and tensor SR state are considerably larger for given $\alpha$, leading to a smaller critical radius for outspiral, but the maximum value of $\alpha$ is also more severely constrained from the depletion time consideration. We estimate the detectability of outspiral in the sensitive band of space-borne GW detectors and also the pulsar-timing detection threshold for general process, finding that the minimum detectable boson mass is lighter for a heavier host BH, and without DF the detectable vector or tensor boson mass is lighter than the scalar case. The inclusion of DF leads to additional detectable parameter space corresponding to inspiral.

Our discussions are focused on circular orbit on the GA's equatorial plane, but as elaborated in Appendix~\ref{Appendix_C}, this model can be extended straightforwardly to more general orbits. For larger orbit eccentricity both $P_{DF}$ and $P_{gw}$ are boosted relative to $P_{DC}$; this has two implications: (i) the critical orbit frequency of outspiral becomes lower, mainly due to the increased $P_{gw}$, (ii) DF becomes more important at given orbit frequency, hence the observable parameter space of DF would be enlarged (though DF is still expected to be insignificant in the $P_{DC}$-dominated phase since it tends to decrease exponentially with the orbit radius). Moreover, for elliptic orbit the cloud's gravity leads to additional effects of orbit precession. We have also neglected the possible matter accretion of the cloud's host BH from the background environment \cite{Akil:2023kym}, which would compete with the mass loss due to cloud mass depletion but may also lead to a larger cloud mass \cite{Hui:2022sri}.

Besides the GW emitted by the binary orbit motion, the GW emitted by the cloud, typically at higher frequency and even larger amplitude (see Fig.~\ref{power_example}) may also be detectable \cite{Siemonsen:2022,Jones:2023fzz}. A joint detection of the binary GW and cloud GW would be a distinctive signature of such systems, and would help to break the degeneracy of the mass change predicted by other scenarios.
Finally, even for the mass change within GA there are still other possibilities; e.g., for complex bosonic fields the cloud may have negligible GW emission \cite{book_superradiance}, while self-interaction of the bosonic fields may lead to additional mass loss \cite{Baryakhtar_2021}, which could enrich  the phenomenology discussed above.

In view of these, we hope to return to this subject in the future with a systematic investigation of the general orbits taking into account the environmental accretion effects, the cloud ionization and possibly other mass loss mechanisms, and the effects on the GW waveforms~\cite{Eda:2013gg, Yue:2017iwc, Yue:2018vtk, Traykova:2021dua, Li:2021pxf, Chung:2021roh, Zhang:2022rfr, Traykova:2023qyv, Bamber:2022pbs, Cole:2022ucw, Su:2021dwz}.

\section*{acknowledgement}
This work is supported by National Key Research and Development Program of China (Grant No.2021YFC2201901), and National Natural Science Foundation of China under Grants No.12147103 and 11851302.

\appendix
\section{Wave Function of Gravitational Atom}\label{Appendix_A}
The orthonormalized wave functions of states $|nljm\rangle$ are given by
\begin{equation}
	\boldsymbol{\Psi}_{nljm}(t,\mathbf{r})=R_{nl}(r)\mathbf{Y}_{ljm}(\theta,\phi)e^{-i(\omega_{nljm}-\mu)t}~,
\end{equation}
where $R_{nl}(r)\equiv r_c^{3/2}R_{nl}(x)$ (with $x\equiv r/r_c$) is the hydrogenic radial function:
\begin{equation}
	R_{n l}(x)= \sqrt{\left(\frac{2}{n}\right)^3 \frac{(n-l-1) !}{2 n(n+l) !}}\left(\frac{2 x}{n}\right)^{l}
	e^{-\frac{x}{n}} L_{n-l-1}^{2 l+1}\left(\frac{2 x}{n}\right)~,
\end{equation}
For scalar fields, the angular function of mode $|nlm\rangle$ is given by the spherical harmonics $Y_{lm}(\theta,\phi)$. For vector field, the angular function of mode $|nljm\rangle$ is given by the purely orbital vector spherical harmonics \cite{Baryakhtar:2017ngi}:
\begin{equation}
Y_{ljm}^i=\sum_{m_l=-l}^{l} \sum_{m_s=-1}^1\left\langle\left(1, m_s\right),\left(l, m_l\right) \mid j, m\right\rangle \xi_i^{m_s} Y_{l m_l}(\theta, \phi)~,
\end{equation}
where $\boldsymbol{\xi}^0=\mathbf{e}_z, \boldsymbol{\xi}^{\pm 1}=\mp \frac{1}{\sqrt{2}}(\mathbf{e}_x \pm i \mathbf{e}_y)$ is a set of orthonormal polarization basis and $\left\langle\left(1, m_s\right),\left(l, m_l\right) \mid j, m\right\rangle$ the Clebsch-Gordan coefficients. For tensor fields, the angular function of mode $|nljm\rangle$ is given by the purely orbital spin-2 tensor spherical harmonics \cite{Brito_2020}:
\begin{align}
Y_{ljm}^{ik}
=\sum_{m_l=-l}^{l} \sum_{m_s=-2}^2\left\langle\left(2, m_s\right),\left(l, m_l\right) \mid j, m\right\rangle t_{ik}^{m_s} Y_{l m_l}(\theta, \phi)~,
\end{align}
with $t_{ik}^{m_s}=\sum_{m_1,m_2=-1}^{1}\langle\left(1, m_1\right),\left(1, m_2\right) \mid 2, m_s\rangle \xi_i^{m_1}\xi_k^{m_2}$. The analytical results for the spectra $\omega_{nl(j)m}$ of spin-0, -1, -2 GA (which are accurate for small $\alpha$) can be found in \cite{B2,Brito_2020}. The velocity of a Schr\"{o}dinger field $\Psi=|\Psi|e^{is}$ is given by $\mathbf{u}=\frac{1}{\mu}\nabla s=\frac{i}{2\mu |\Psi|^2}(\Psi\nabla\Psi^*-\Psi^*\nabla\Psi)$. For scalar state, $s\propto m\phi$, so $\mathbf{u}=\frac{m}{\mu r\sin\theta}\mathbf{e}_\phi$. For vector and tensor state with azimuthal quantum number $m$, the field components $\Psi_i\propto e^{i m_i\phi}$ generally have distinct values of $m_i$.

In the nonrelativistic Newtonian limit, the system can be described by the Lagrangian (neglecting self-gravity)
\begin{align}
\begin{autobreak}
\mathcal{L}=
\frac{M_c}{\mu}\text{Tr}\left[\frac{1}{2}\left(i \boldsymbol{\Psi}^\dagger \dot{\boldsymbol{\Psi}}+\text {c.c.}\right)-\frac{1}{2\mu} \nabla \boldsymbol{\Psi}^\dagger \cdot \nabla \boldsymbol{\Psi}+\frac{\alpha}{r}|\boldsymbol{\Psi}|^2\right],
\end{autobreak}
\end{align}
which leads to the Schr\"{o}dinger equation for each field component. But the energy-momentum tensor should be obtained from the relativistic Lagrangian. The equations of motion for massive scalar, vector and spin-2 field read
\begin{equation}
\begin{aligned}
\square \phi &=\mu^2\phi~,
\\
\square A_b-R_{cb}A^c&=\mu^2 A_b~,
\\
\square H_{ab}+2R_{acbd}H^{cd}&=\mu^2 H_{ab}~,
\end{aligned}
\end{equation}
with $A^b_{;b}=0$ and $H^{ab}_{;a}={H^a}_a=0$, in the nonrelativistic limit $A^0,H^{0b}\approx 0$. For the scalar field, the nonrelativistic ansatz corresponds to
\begin{equation}
\phi=\sqrt{\frac{M_c}{\mu}}\frac{1}{\sqrt{2\mu}}(\Psi e^{-i\mu t}+\text{c.c.})
\end{equation}
for the Proca field,
\begin{equation}
A_i=\sqrt{\frac{M_c}{\mu}}\frac{1}{\sqrt{2\mu}}(\Psi_i e^{-i\mu t}+\text{c.c.})
\end{equation}
and for the massive spin-2 field,
\begin{equation}
H_{ij}=\sqrt{\frac{M_c}{\mu}}\frac{1}{\sqrt{2\mu}}(\Psi_{ij} e^{-i\mu t}+\text{c.c.})
\end{equation}
[corresponding to the Fierz-Pauli Lagrangian with mass term $\frac{1}{2}\mu^2(H^2-H_{ab}H^{ab})$]. In the non-relativistic limit and in a nearly flat spacetime background, the energy density is then given by $\rho=T_{00}=M_c \text{Tr}(\boldsymbol{\Psi}\boldsymbol{\Psi}^*)=\frac{M_c}{r_c^3}g$, with $g=R_{nl}^2(x)\text{Tr}(Y_{ljm}Y^*_{ljm})$.  Due to axisymmetry, the mass quadrupole moment of the cloud is given by $\bar I_{ij}=Q_c(\hat z_i\hat z_j-\frac{1}{3}\delta_{ij})$ where $\hat{\mathbf{z}}$ is the direction of BH spin, with
\begin{equation}
Q_c=\int d^3 r \rho(\mathbf{r}) r^2 P_2(\cos \theta).
\end{equation}
For scalar 211 state (same as vector 2122 and tensor 2133 state), $g=\frac{1}{64}x^2e^{-x}\sin^2\theta$ and $Q_c=-6M_cr_c^2$. For scalar 322 state, $g=\frac{x^4 e^{-\frac{2 x}{3}}\sin ^4\theta}{26244 \pi }$ and $Q_c=-36M_cr_c^2$. For the scalar 433 state, $g=\frac{e^{-\frac{x}{2}} x^6 \sin ^6\theta}{37748736 \pi}$ and $Q_c=-120M_cr_c^2$. For vector 1011 state (same as tensor 1022 state), $g=\frac{1}{\pi}e^{-2x}$ and $Q_c=0$. For tensor 2111 state, $g=\frac{13-\cos 2\theta}{1280\pi}x^2e^{-x}$ and $Q_c=-\frac{3}{5}M_cr_c^2$.

Since the spatial gradients of the field are nonrelativistically suppressed, the other components of the energy-momentum tensor are approximately given by $T_{i0}=0$,
\begin{equation}
\begin{aligned}
T_{ij} &=\frac{\delta_{ij}}{2}\left[(\dot\phi)^2-\mu^2\phi^2\right],
\\
T_{ij} &=\frac{\delta_{ij}}{2}\left[(\dot A_k)^2-\mu^2(A_k)^2\right]-\left[\dot A_i\dot A_j-\mu^2A_iA_j\right],
\\
T_{ij} &=\frac{\delta_{ij}}{2}\left[(\dot H_{kl})^2-\mu^2(H_{kl})^2\right]-2\left[\dot H_{ik}\dot H_{kj}-\mu^2H_{ik}H_{kj}\right],
\end{aligned}
\end{equation}
for scalar, Proca and spin-2 field, respectively. These components oscillate at frequency $2\mu$ and would source metric perturbations oscillating at the same frequency. Note that the binary orbit frequency $\Omega=n\mu$ corresponds to the orbit radius $x=[(1+q)\alpha^4/n^2]^{1/3}$, which for $\alpha\ll 1$ is much smaller than the Bohr radius; hence the possible orbit resonance from these oscillating metric perturbations is irrelevant in the perturbative regime of GA.

\section{Correction to $M_c$ and $P_{gw}$}\label{Appendix_B}
\begin{figure}[h]
	\centering
	\includegraphics[width=0.5\textwidth]{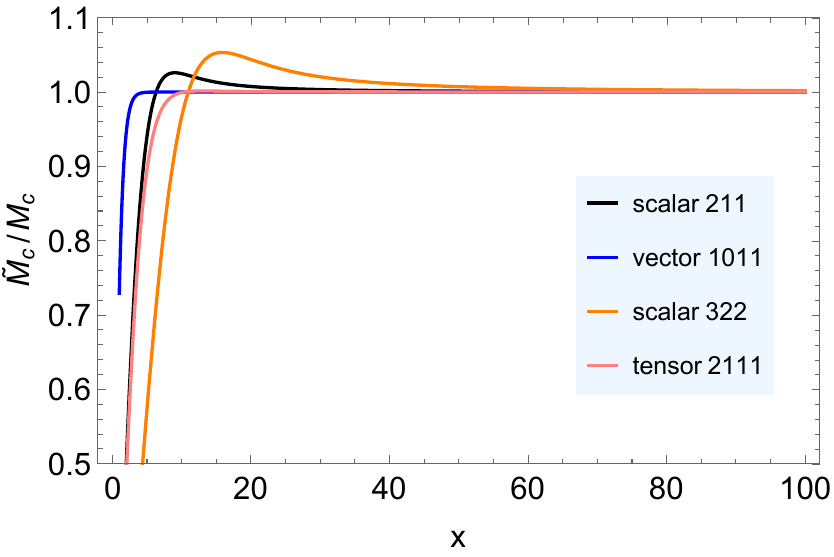}\quad
	\caption{Effective cloud mass experienced by the companion.}\label{Mct}
\end{figure}

The effect of mass depletion on the binary dynamics originates from the cloud's gravitational force on the companion. As leading order approximation we consider only the Newtonian potential of the cloud in a flat background (since the other linear metric perturbations are already negligible for $x\gtrsim 10$). Due to axisymmetry, the Newtonian potential sourced by the cloud can be expanded in terms of the Legendre polynomials:
\begin{equation}
\Phi(r, \theta)=\sum_{n=0}^\infty \Phi_n(r) P_n(\cos \theta)~,
\end{equation}
where
\begin{equation}
\begin{aligned}
\Phi_n(r)=&-\frac{2 \pi}{(n+1 / 2) r^{n+1}} \int_0^r (r')^{n+2} \rho_n\left(r^{\prime}\right) d r^{\prime}
\\
&-\frac{2 \pi r^n}{n+1 / 2} \int_r^{\infty} (r')^{1-n} \rho_n\left(r^{\prime}\right) d r^{\prime}~,
\end{aligned}
\end{equation}
and 
\begin{equation}
\rho_n(r)=(n+1 / 2) \int_0^\pi \rho(r, \theta) P_n(\cos \theta) \sin \theta d \theta~.
\end{equation}
Including the cloud's gravity, the binary potential energy is
\begin{equation}
E_p=M_*\Phi-\frac{M_*M}{r}~,
\end{equation}
so we can define an effective cloud mass\footnote{Here we are using the local approximation $\Phi r\approx \text{const}$. Another approximation could be $\tilde M_c=r^2\mathbf{e}_r\cdot \nabla\Phi$. Since we assume a Keplerian orbit in the leading order, $\tilde M_c-M_c$ is treated as a small perturbation, so its contribution to the binary orbital energy is negligible (otherwise even for the equatorial plane circular orbit we need to use a modified Keplerian relation). $\tilde M_c$ then mainly affects $P_{DC}$, the difference between these two approximations is significant only at very small orbit radius, where the DC effect is negligible.}
\begin{equation}
\tilde M_c=-\Phi r~.
\end{equation}
For $\theta=\pi/2$, $\Phi(r,\theta)=\Phi(r)$, the results for several states are depicted in Fig.~\ref{Mct}. It can be seen that the deviation of $\tilde M_c$ from $M_c$ is small. Interestingly, for a nonspherical state $\tilde M_c/M_c$ is not a monotonic function of radius.

Since $\tilde M_c$ depends on radius, $\frac{d}{dt}\tilde M_c=\frac{\tilde M_c}{M_c}\dot M_c+(\frac{\tilde M_c}{M_c})_{,x}\dot x$, but the second contribution is negligible since it is a second-order effect due to orbit evolution. From the mass change of the cloud, the binary GW radiation power $P_{gw}=\frac{1}{5} {\dddot{\bar I}}_{ij}{\dddot{\bar I}_{ij}}$ receives a correction via the change of binary mass quadrupole\footnote{The use of quadrupole formula even in the case of time-varying mass is not strictly justified, but this problem is irrelevant for the very slow mass depletion considered here.} $I_{ij}=\frac{M_*}{1+q}x_ix_j$ with $q=\frac{M_*}{M+\tilde M_c}$, but we have checked that it is completely negligible for any reasonable parameters. 

The gravity of the cloud could also affect the binary's GW emission; this is particularly relevant for an extreme-mas-ratio system with sufficiently small orbit radius or large cloud mass, which has been recently investigated in \cite{PhysRevD.108.084019} for a circular binary around a complex scalar cloud.

\section{General Orbit}\label{Appendix_C}
In this appendix we outline the Newtonian analysis for a general inclined elliptical orbit, neglecting the possible precession of the orbit plane. In the untilted BH-centered frame $(x,y,z)$ with the $z$ axis parallel to the BH spin, the companion's coordinate position is $(r,\theta,\phi)$, and we denote its position in the tilted BH-centered frame $(X,Y,Z)$ with rotated $X$ axis and $Z$ axis parallel to the orbit normal to be $(r,i,\varphi+\varphi_0)$, where $i$ is the orbit plane's inclination angle (relative to the BH's equatorial plane), $\varphi$ the true anomaly, and $\varphi_0$ the longitude of the periastron on the orbit plane. The two sets of coordinates are related by $\cos\theta=\sin i\cos(\varphi_0+\varphi)$, $\sin\theta=\sqrt{\sin^2(\varphi_0+\varphi)+\cos^2i\cos^2(\varphi_0+\varphi)}$, and $\tan \phi=\sec i\tan (\varphi_0+\varphi)$. The cloud's density distribution in the orbit plane is then given by $\rho(r,\theta(\varphi))$.

For elliptical orbit, the convenient parametrization is
\begin{equation}
\begin{aligned}
	r=\frac{a(1-e^2)}{1+e\cos\varphi}=a(1-e\cos z),
	\quad
	z-e\sin z=\Omega t,
\end{aligned}
\end{equation}
where $a$ is the semimajor axis, $e$ the eccentricity, and $z\in [0,2\pi]$ the eccentricity anomaly. The orbit velocity, energy and angular momentum are given, respectively, by
\begin{align}
v&=a\Omega\sqrt{\frac{1+e\cos z}{1-e\cos z}}~,
\\
E/\mu&=-\frac{M_{tot}}{2a}~,
\\
L/\mu&=\sqrt{M_{tot}a(1-e^2)}~,
\end{align}
with $M_{tot}=M_1+M_2$, $\mu=M_1M_2/M_{tot}$ (in this appendix $\mu$ refers to the binary's reduced mass, instead of the boson mass), and the Kepler relation $\Omega=\sqrt{\frac{M_{tot}}{a^3}}$. The radial and azimuthal velocity are, respectively,
\begin{equation}
v_r=\sqrt{\frac{M_{tot}}{a(1-e^2)}}\,e\sin\varphi,\quad
v_\varphi=\frac{\sqrt{M_{tot}a(1-e^2)}}{r}.
\end{equation}
 
In the absence of mass variation, the orbit evolves according to
\begin{equation}
\begin{aligned}
-\dot E =P_{gw}+P_{DF}
,\quad
\dot L =(\dot L)_{gw}+(\dot L)_{DF},
\end{aligned}
\end{equation}
leading to evolution of the osculating orbit elements $a$ and $e$. For a perturbing (relative) acceleration of the two-body $\mathbf{F}=F_\varphi \mathbf{e}_\varphi+F_{r}\mathbf{e}_r+F_z\mathbf{e}_z$ (where $\mathbf{e}_z$ is perpendicular to the orbit plane) acted on the system, the orbit evolution is given by \cite{danby1992fundamentals}
\begin{subequations}\label{secular}
\begin{align}
	\frac{\dot{a}}{a}&=\frac{2}{\Omega}\left\{\frac{e \sin \varphi}{a \sqrt{1-e^2}} F_r+\frac{\sqrt{1-e^2}}{r} F_\varphi\right\},
	\\
	\dot{e}&=\frac{\sqrt{1-e^2}}{a \Omega}\left\{(\cos \varphi+\cos z) F_\varphi+\sin \varphi F_r\right\},
	\\
	\dot\varphi_0&=\frac{\sqrt{1-e^2}}{a e \Omega}\left\{\left[1+\frac{r}{a (1-e^2)}\right] \sin \varphi F_\varphi-\cos \varphi F_r\right\}.
\end{align}
\end{subequations}

\subsection{Dynamical Friction} 
\begin{figure*}[hbt]
	\centering
	\includegraphics[width=0.39\textwidth]{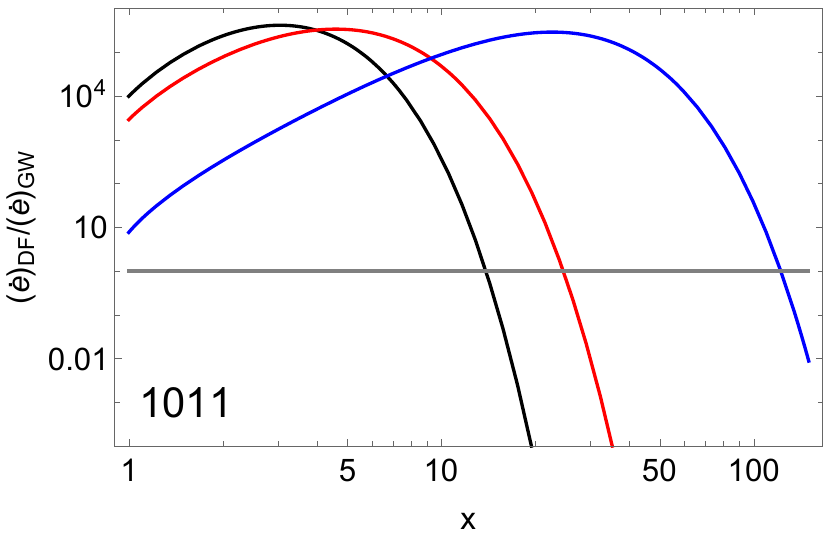}
 \quad
    \includegraphics[width=0.57\textwidth]{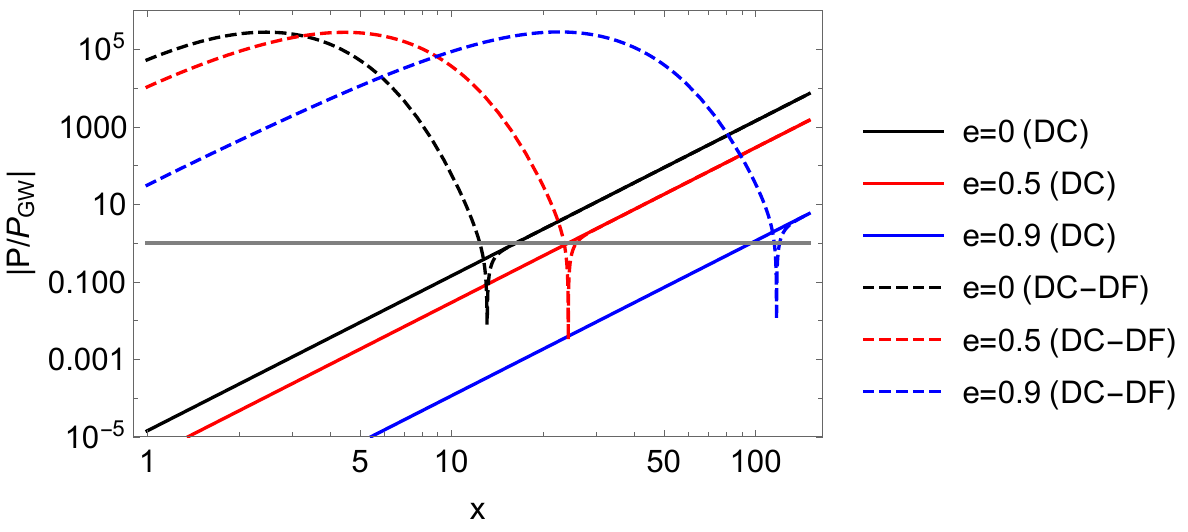}
	\caption{Changing rate of eccentricity (left) and effective power (right) versus $x=a/r_c$, for saturated vector 1011 state with $q=0.01$, $\alpha=0.03$, and $\beta=0.01$, using the DF model \eqref{DF_model}. In the right figure, the dashed lines correspond to $P_{DC}-P_{DF}$ and the solid lines correspond to $P_{DC}$. For $P_{DC}$, we use for simplicity $\tilde M_c=M_c$; the relative difference is within $\mathcal{O}(1)$ for $x>1$.}\label{1011_elliptic}
\end{figure*}

For a general friction force parallel to the velocity, $\mu\mathbf{F}=\mathcal{F} \mathbf{v}/v$, the secular evolution is
\begin{subequations}
\begin{align}
\left\langle\frac{\dot a}{a}\right\rangle&=\frac{1}{2\pi}\int_0^{2\pi}dz\, 
\frac{\mathcal{F}}{\mu}\sqrt{\frac{a}{M_{tot}}} 2\sqrt{1-e^2\cos^2 z},
\\
\left\langle \dot e \right\rangle&=\frac{1}{2\pi}\int_0^{2\pi}dz\, 
\frac{\mathcal{F}}{\mu}\sqrt{\frac{a}{M_{tot}}} \frac{2(1-e^2)(\cos z-e\cos^2 z)}{\sqrt{1-e^2\cos^2 z}},
\end{align}
\end{subequations}
while for $\mathcal{F}(\varphi)=\mathcal{F}(-\varphi)$, $\langle\dot\varphi_0\rangle=0$; hence the dynamical friction does not contribute to the periastron shift if $\rho(\varphi)=\rho(-\varphi)$, e.g., for $i=0$. For elliptic orbit we define the effective power by $\dot x_a=-\frac{2}{qM\alpha^2}P(x_a)x_a^2$ with $x_a=a/r_c$. The generalization of DF power \eqref{p_df} to elliptical orbit is
\begin{equation}\label{DF_e_ratio}
\frac{P_{DF}(e,i)}{P_{DF}(e=0,i=0)}=\frac{\int_0^{2\pi}dz\,xC_\Lambda(x)\sqrt{1-e^2\cos^2z}\,g(x,\theta)}{2\pi x_aC_\Lambda(x_a)\,g(x_a,\pi/2)}.
\end{equation}
For $P_{DF}$, orbit inclination relative to the equatorial plane leads to reduced cloud mass density along the orbit for nonspherical states, we consider in the following $i=0$. A larger eccentricity, on the other hand, means that the companion dives into a denser region of the cloud and hence $P_{DF}$ could increase. For high eccentricity or large orbit radius, the ratio \eqref{DF_e_ratio} is significant and we find that $P_{DF}(e)$ is mainly determined by $g(x_a(1-e))$, i.e., the cloud density at the periastron. Similarly, the accretion effect is proportional to the cloud density and subjects to the same suppression as the circular case (note that accretion is irrelevant if the companion is not a BH). Since the DF decreases exponentially at large radius, it is still not important for the outspiral phase; however, the binary GW power $P_{gw}(x_a,e)=P_{gw}(x_a,e=0)f(e)$ is boosted by the Peters-Mathews factor $f(e)=\frac{37 e^4/96+73 e^2/24+1}{\left(1-e^2\right)^{7/2}}$, pushing the outspiral phase to a lower orbit frequency. At small radius, the DF could significantly modify the eccentricity evolution; the same as the effective power we find that $\frac{\langle \dot e\rangle _{DF}}{\langle \dot e\rangle _{gw}}\propto \beta\alpha^{-5}$, where $\langle \dot e\rangle_{gw}$ is the contribution from binary GW emission. At large radius, $\langle\dot e\rangle=\langle\dot e\rangle_{gw}$, which could nevertheless give some corrections to the result of secular evolution for circular orbit discussed in the main text. As a concrete example we show the effective powers together with the changing rate of eccentricity for the vector 1011 state in Fig.~\ref{1011_elliptic}.

\subsection{Cloud Mass Depletion} For a binary with time-varying mass, in the absence of dissipation, $\mu\ddot{\mathbf{r}}=-\frac{M_1(t)M_2(t)}{r^3}\mathbf{r}$; hence $L/\mu=r^2\dot\varphi$ is conserved, for circular orbit this leads to $\dot r=-\frac{\dot M_{tot}}{M_{tot}}r$. A more careful treatment \cite{hadjidemetriou1963two,dosopoulou2016orbital1,dosopoulou2016orbital2} assuming the mass loss is isotropic (i.e., the mass change itself does not carry away linear momentum) would show that the mass change manifests as a perturbing acceleration $\mathbf{F}_{DC}=-\frac{1}{2}\frac{\dot M_{tot}}{M_{tot}}\mathbf{v}$, and its contribution to the effective power reads $P_{DC}=\frac{q\alpha^2}{2(1+q)x_a}\dot M_1$. For circular orbit with $\dot M_2=0$ this corresponds to the average loss rate of orbital angular momentum:
\begin{equation}
(\dot L)_{DC}=
\frac{M_2^2}{M_{tot}^{3/2}}\dot M_1\sqrt{r}~,
\end{equation}
which is negligible only for $q\ll 1$. In comparison, if $L$ remains constant during the process of mass change, the resulted effective power is
\begin{equation}\label{mass_transfer}
P(x)=\frac{q\alpha^2}{2(1+q)x}[\dot M_1(1+2q)+\dot M_2(1+2q^{-1})]~.
\end{equation}

As seen from Eq.~\eqref{secular}, the isotropic mass change (averaged over time) affects neither the eccentricity evolution nor the periastron precession. But for inclined elliptical orbit, the gravity of the cloud can lead to both periastron and orbit plane precession (for a non-spherical state such as scalar 211), unrelated to the cloud mass depletion.

\subsection{Ionization and Resonance} Finally we briefly discuss how to incorporate the mass depletion in the ionization and resonance dynamics. If we identify the DF with ionization, the orbit dynamics can be set up using the ionization model of \cite{B4}, but the orbit eccentricity and inclination introduce additional complexities \cite{Takahashi:2023flk}. For simplicity, we consider only the equatorial plane circular orbit and the ionization of a single bound mode. The cloud mass changes as $\dot M_c=-P_{gw,c}-\dot M_*-M_c\sum_{g}\left[\frac{\mu |\eta|^2}{k}\Theta(k^2)\right]_g$, where $\eta$ is the mixing matrix element between the single bound state being ionized and the continuum state with wave number $k^{(g)}$ and azimuthal quantum number $m'=m+g$ satisfying $\frac{(k^{(g)})^2}{2\mu}=-\mu\frac{\alpha^2}{2n^2}\pm g\Omega$ ($\pm$ stands for corotating or counterrotating orbit; here we adopt $\Omega>0$), the orbit angular momentum changes due to ionization according to $(\dot L)_{ion}=\mp M_c\sum_g(m+g)\left[\frac{|\eta|^2}{k}\Theta(k^2)\right]_g$. Together with the cloud's angular momentum $S_c=\frac{m M_c}{\mu}$, and using the angular momentum balance $\dot L\pm (\dot S_c+\frac{m}{\mu}P_{gw,c})=-\frac{P_{gw}}{\Omega}+(\dot L)_{ion}+(\dot L)_{DC}$ (assuming the accretion process giving rise to $\dot M_*$ does not change the total angular momentum), the contribution to the effective power from ionization and $\dot M_*$ is \cite{B4}
\begin{equation}
\begin{aligned}
P_{ion}(x) =& M_c\Omega\sum_g g\left[\frac{|\eta|^2}{k}\Theta(k^2)\right]_g 
\\
&+\frac{M\Omega}{\alpha} \dot M_* \left[\frac{(2+q)\sqrt{x}}{2(1+q)^{3/2}}\mp m\right]~,
\end{aligned}
\end{equation}
The first term in the second line is same as Eq.~\eqref{mass_transfer} for $\dot M_1=0$, since here the contribution of cloud mass to the binary orbital energy is neglected. For small mass ratio and large orbit radius, the second line is nothing but $\dot M_* v^2$, see also Sec.~\ref{accretion_compare}.

For the orbit evolution during resonance (again restricting to equatorial plane circular orbit and neglecting the ionization and accretion), from angular momentum balance the orbit evolution is given by
\begin{equation}
\dot{x}=(\dot x)_{others}\mp 2(1+q)^{1 / 2} q^{-1} \alpha x^{1 / 2} M^{-2}([\dot S_c]_\text{eff}+\dot J)
\end{equation}
(where the contribution of $\dot M$ to $L$ has been neglected). Here $(\dot r)_{others}$ includes all effects other than the cloud-orbit angular momentum exchange, $J=M^2\chi$ is the BH spin, the angular momentum of the cloud $S_c=\frac{M_c}{\mu} \sum_i m_i\left|c_i\right|^2$ evolves according to $i \dot{c}_i=H_{i j} c_j$, with the mixing Hamiltonian $H_{ij}=\langle \boldsymbol{\Psi}_i|V_*|\boldsymbol{\Psi}_j\rangle+\omega^{(i)}\delta_{ij}$. $[\dot S_c]_\text{eff}$ is $\dot S_c$ with the contribution from cloud depletion removed. The mass and angular momentum conservation of the SR process imply that $\dot M+\sum_i 2\omega_I^{(i)}M_c|c_i|^2=0$ and $\dot J+\sum_i 2\omega_I^{(i)}\frac{M_c}{\mu}m_i|c_i|^2=0$. For hyperfine transitions the variation of BH mass and spin can be important for the mode evolution \cite{Takahashi:2023flk}. Due to its long timescale, the cloud mass depletion is not expected to play any roles in the ``quantum dynamics'' of $c_i$ which is mainly driven by the oscillating gravitational perturbation, but its orbit effect through $P_{DC}$ may still be relevant; e.g., for a sufficiently large cloud mass, the orbit evolution before the resonance would be modified.

%

\bibliography{main}
\end{document}